\documentclass{emulateapj}
\usepackage{graphicx}
\usepackage[usenames,dvipsnames]{color}

\providecommand{\sorthelp}[1]{}

\shorttitle{Magnetic fields at the infrared dark cloud G14.225-0.506}
\shortauthors{Santos et al.}

\begin{document}

\title{
Magnetically Dominated Parallel Interstellar Filaments at the \\ Infrared Dark Cloud G14.225-0.506
\thanks{Based on observations collected at Observat\'orio
 do Pico dos Dias, operated by Laborat\'orio Nacional de Astrof\'\i sica 
 (LNA/MCT, Brazil).}}

\author{F\'abio P. Santos\altaffilmark{1}, 
Gemma Busquet\altaffilmark{2},
Gabriel A. P. Franco\altaffilmark{3}, 
Josep Miquel Girart\altaffilmark{2,4} \&
Qizhou Zhang\altaffilmark{4}
}

\affil{\altaffilmark{1}  Department of Physics and Astronomy, Northwestern University, 2145 Sheridan Road, \\ 
Evanston, IL 60208, USA; fabiops@northwestern.edu}
\affil{\altaffilmark{2} Institut de Ci\`encies de l'Espai (CSIC-IEEC), Campus UAB, Carrer de Can Magrans, S/N E-08193 \\ 
Bellaterra, Catalunya, Spain; busquet@ice.cat, girart@ice.cat}
\affil{\altaffilmark{3} Departamento de F\'isica -- ICEx -- UFMG, Caixa Postal 702, 30.123-970 \\
Belo Horizote, MG, Brazil; franco@fisica.ufmg.br}
\affil{\altaffilmark{4} Harvard-Smithsonian Center for Astrophysics, 60, Garden Street, \\ 
Cambridge, MA 02138, USA; qzhang@cfa.harvard.edu}

\begin{abstract} 
The G14.225-0.506 infrared dark cloud (IRDC G14.2) displays a remarkable complex of 
parallel dense molecular filaments projected on the plane of the sky. Previous dust emission 
and molecular-line studies have speculated whether magnetic fields
could have played an important role in the formation of such long-shaped structures, 
which are hosts to numerous young stellar sources. In this work we have conducted
a vast polarimetric survey at optical and near-infrared wavelengths in order to study the 
morphology of magnetic field lines in IRDC G14.2 through the observation of background stars.
The orientation of interstellar polarization, which traces magnetic field lines, is 
perpendicular to most of the filamentary features within the cloud. Additionally, the larger-scale 
molecular cloud as a whole exhibits an elongated shape also perpendicular to magnetic fields.
Estimates of magnetic field strengths indicate values in the range $320 - 550\,\mu$G, 
which allows sub-alfv\'enic conditions, but does not prevent the gravitational collapse of hub-filament structures,
which in general are  close to the critical state.
These characteristics suggest that magnetic fields played the main role in regulating the 
collapse from large to small scales, leading to the formation of series of parallel elongated structures.
The morphology is also consistent with numerical simulations that show how gravitational
instabilities develop under strong magnetic fields. 
Finally, the results corroborate the hypothesis that a strong support from internal magnetic fields
might explain why the cloud seems to be contracting on a time scale $2-3$ times larger than 
what is expected from a free-fall collapse.
\end{abstract}


\section{Introduction}
\label{introduction}

Filamentary structures in the interstellar medium (ISM) are commonly 
observed in many different types of environments, such as diffuse nearby 
clouds \citep{penprase1998,mcclure2006}, giant molecular clouds \citep{lis1998,hill2011}, 
H{\sc ii} regions \citep{anderson2012,minier2013}, and supernova remnants \citep{gomez2012}.
Their presence in the Milky Way Galaxy has typically been revealed by numerous 
different observing techniques, including visual extinction, HI emission, 
molecular line surveys and dust thermal emission.
In particular, dense molecular filaments are in general associated with star-forming regions.
\citet{myers2009} pointed out, for instance, that all the nearest low-mass star formation 
sites (within $300\,$pc from the Sun) seem to present a hub-filament structure, 
with some of them showing evenly-spaced parallel filaments.

Although filaments are known for many decades, during more recent years, observations of dust thermal emission 
from the {\it Herschel} space observatory \citep{pilbratt2010} provided a groundbreaking 
understanding of filaments in the ISM, showing that they are in fact
ubiquitous especially within giant molecular clouds \citep{andre2010,molinari2010}, 
which includes both quiescent and star-forming regions.
The recognition of filaments as active sites of star-formation was made clear
by the fact that most of the observed pre-stellar cores seem to form in gravitationally unstable filaments 
\citep{andre2010,arzou2011}. That led to an increasing interest in explaining how these 
structures are formed and how they evolve.
Although turbulent motions in the ISM might be responsible for the formation of some filaments \citep{arzou2011}, 
other plausible explanations are the convergence of flows, large-scale collisions between 
filaments, or gravitational instabilities \citep{schneider2010,jimenez2010,nakamura2012,nakajima1996,vanloo2014}, a scenario which is also supported by numerical models \citep{gomez2014}.

In addition, it is well known that the interstellar medium is entirely threaded by a large-scale
structure of magnetic field lines that pervades the whole Galaxy \citep{mathewson1970,reiz1998,heiles2000,santos2011}. This includes the filaments as well as the dense molecular cores where the star formation is taking place \citep[e.g.,][]{alves2008,girart2006,girart2009,zhang2014,li2015}. 
In general, a small level of ionization is sufficient to provide enough coupling between the magnetic fields
and the interstellar gas \citep{heiles2005}.
Indeed, magnetic fields 
might also play an important role in generating filamentary structures, as suggested by several authors \citep{nagai1998,nakamura2008,li2013,li2015}.

G14.225-0.506 (hereafter IRDC G14.2) is an infrared dark cloud at a distance of 
$1.98^{+0.13}_{-0.12}\,$kpc \citep{xu2011, wu2014} that shows an intricate pattern of filaments.
These filaments are clearly seen in absorption against the bright mid-infrared background Galactic emission, 
as identified by \citet{2009peretto} using {\it Spitzer Space Telescope} data\footnote{Based 
on the Galactic Legacy Infrared Mid-Plane Survey Extraordinaire \citep[GLIMPSE,][]{2003benj} 
and the Multiband Imaging Photometer for Spitzer Galactic Plane Survey \citep[MIPSGAL,][]{2009carey}.}.
This region is part of a larger complex of clouds including 
the well-known M17 star-forming area \citep{elm1976}. 
Later studies revealed  
star formation signs such as H$_{2}$O maser emission \citep{1981jaffe,palagi1993,wang2006} and 
emission from dense gas tracers toward IRAS 18153-1651,
which is one of the bright infrared sources in the region \citep{plume1992,anglada1996}, with a luminosity of $1.1\times10^4~L_{\sun}$. 
Furthermore, many young stellar objects were later identified by \citet[][who labeled this region 
as M17 SWex]{povich2010}, including several Class 0 and I sources. Although IRDC G14.2 does
not appear to host very massive stars, a few ultra-compact H{\sc ii} regions 
are located amongst its filamentary structures \citep{bronfman1996,jaffe1982}.
\citet[][herafter Paper I]{busquet2013} presented ammonia observations in IRDC G14.2, inferring that the  
parallel arrangement 
of most filaments could be explained 
by the gravitational collapse of an unstable thin layer threaded by magnetic fields \citep{vanloo2014}.

The sky-projected morphology of magnetic field lines may be mapped through 
studies of the interstellar polarization due to magnetically aligned dust particles, 
either through observations of background starlight, or direct thermal emission from dust.
Although the detailed aspects of the alignment mechanism is one 
of the most long-standing issues in the physics of the ISM, it is now generally 
believed that radiative torques are a dominant effect 
\citep{dolginov1976,draine_wein_1996,lazarian2007}, 
as suggested by different studies 
\citep{whittet2008,andersson2011,alves2014,jones2015}.
Even though different large-scale polarization emission surveys have been 
providing and unprecedented view of magnetic fields in the ISM 
(such as Planck -- \citealt{planck2014} -- and BLASTPol -- \citealt{2016fissel}),
the spatial resolution needed to distinguish filamentary features at distant clouds is still a
challenge, making optical and near-infrared (NIR) polarimetry of background starlight a viable option.

IRDC G14.2 is an ideal target for investigating the role of magnetic fields in generating 
filamentary structures. In this 
work, we present a vast extension of a preliminary polarimetric dataset previously 
shown in Paper I. This includes optical and NIR observations encompassing 
all the filamentary network of IRDC G14.2, as well as the associated large-scale molecular cloud.
In Section \ref{s:obsdata} we describe the polarimetric observations, as well as the data processing. 
Results and analysis are shown in Section \ref{s:results}, which includes 
studies of the relative orientation between magnetic fields with the cloud and internal filaments, 
as well as estimates of various important physical paramenters. A detailed discussion of the results 
is given in Section \ref{s:discussion} and the final conclusions in Section \ref{s:conclusions}.

\section{Observational data}
\label{s:obsdata}

The polarization data used in this work was collected at the 1.6 m telescope 
from the Pico dos Dias Observatory (OPD\footnote{The 
Pico dos Dias Observatory is operated by the Brazilian National Laboratory for 
Astrophysics (LNA), a research institute of the Ministry of Science, Technology and Innovation (MCTI).}, 
Brazil), in a series of observations during July 2011, May 2013 and April 2014.
A small portion of the NIR data, focused on a fraction of the IRDC G14.2 filamentary complex, 
was previously shown in its preliminary version in Paper I.
The current work presents a widely extended version of the same NIR dataset, covering the entire group of 
interstellar filaments (in the H band). Additionally, an optical survey was conducted (using the 
R band), to map an even larger area comprising the large-scale cloud in which the dense filaments are embedded.

The instrumental set was composed by the IAGPOL polarimeter, 
together with an imaging detector, which could be either an optical CCD or a NIR detector 
(HAWAII 1024$\times$1024 pixels -- CamIV), 
depending on the spectral ranges used at each observing run. The polarimeter consists of a 
rotating achromatic half-wave plate followed by an analyzer and a spectral band filter 
\citep[for more information on the instrument and the data reduction process, see][]{magalhaes1996,santos2012}. By rotating 
the half-wave plate in discrete and successive angles of $\psi=22.5\degr$, the 
linear polarization orientation of the incident light changes in steps of $45\degr$. 
The analyzer splits the light beam in two orthogonally polarized components, 
which are simultaneously collected by the detector. The consecutive rotations of the 
half-wave plate produce relative intensity variations between the two components, 
defining an oscillating modulation function proportional to cos$4\psi_i$. 
The flux-normalized $Q$ and $U$ Stokes parameters are determined through a least-square
fitting of this function, using the relative intensity for all targets at each half-wave plate position.
Thereafter, this allows the calculation of the polarization degree ($p$) and orientation in the plane of the sky ($\theta$). 

In this way, $p$ and $\theta$ are found for the majority of the point-like sources 
detected in each observing field. The optical field-of-view covers a $11' \times 11'$
area as opposed to $4' \times 4'$ for the NIR detector. Therefore, a mosaic-mapping 
was adopted to cover a wider area of the sky. For IRDC G14.2,
the R-band observations consisted of a $5 \times 5$ mosaic grid (25 fields), 
resulting in a mapping of a $\sim 53' \times 53'$ area. The H-band observations 
consisted of 8 fields, and were focused on the filamentary structures located 
approximately at the center of the larger-scale area covered by the R-band survey.
For each optical field, two sets of 8 half-wave plate positions were used, 
with a long ($60\,$s) and a short exposure ($10\,$s) at each position.
For the NIR, sixty $10\,$s images where acquired for each half-wave plate 
position, while dithering the telescope to remove the thermal background signal, 
summing up to a total exposure of $600\,$s per half-wave plate position (the 
same procedure was repeated in each field-of-view).

Image processing and photometry were performed using IRAF\footnote{IRAF is distributed by the 
National Optical Astronomy Observatories, which are operated by the Association of 
Universities for Research in Astronomy, Inc., under cooperative agreement with the National 
Science Foundation.} routines \citep{tody1986}, which typically consist of a 
correction of the bias and flat-field pattern, background sky subtraction, detection 
of point-like sources (with a threshold of $5\sigma$ above the local background), flux measurements and configuration
of the image's astrometry (world coordinate system).
Computation of linear polarization for each star was done with the PCCDPACK set 
of routines \citep{pereyra2000}, and calibration of the zero-point polarization 
angle was based on polarimetric standard sources observed each night 
\citep{wilking1980,wilking1982,clemens1990,turnshek1990,larson1996}. 
Finally, de-biased polarization values were computed 
\citep[$p \rightarrow (p^{2}-\sigma_{p}^{2})^{1/2}$, ][]{wardle1974}.
In the analysis that follows, we use only detections with values of $p/\sigma_{p}$ greater
than 4 and 5 for the R and H band samples, respectively.

\section{Results and Analysis}
\label{s:results}

\subsection{Polarization maps and general interstellar features}

\begin{figure*}
   \centering
   \includegraphics[width=\textwidth]{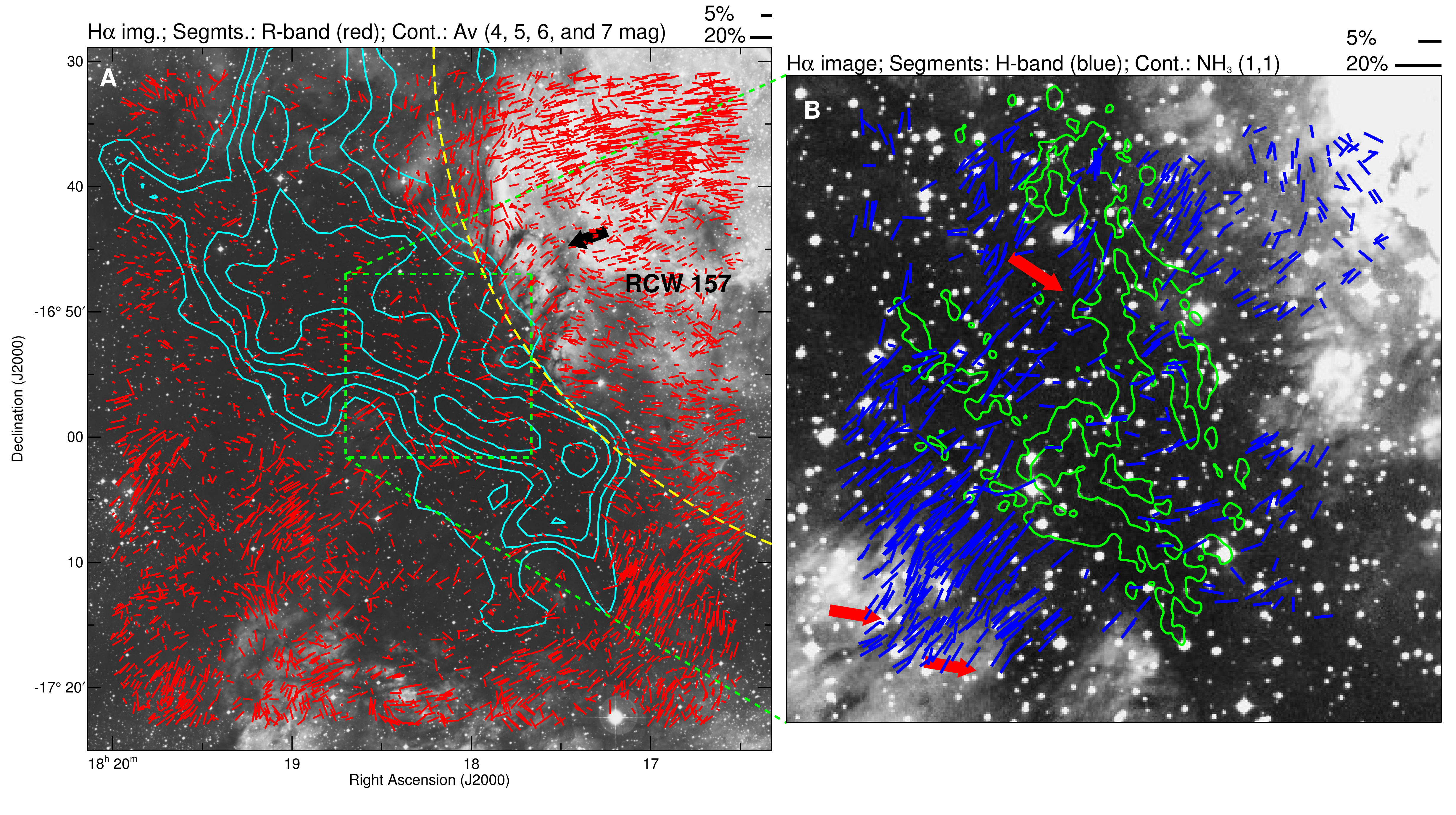} \\
   \includegraphics[width=\textwidth]{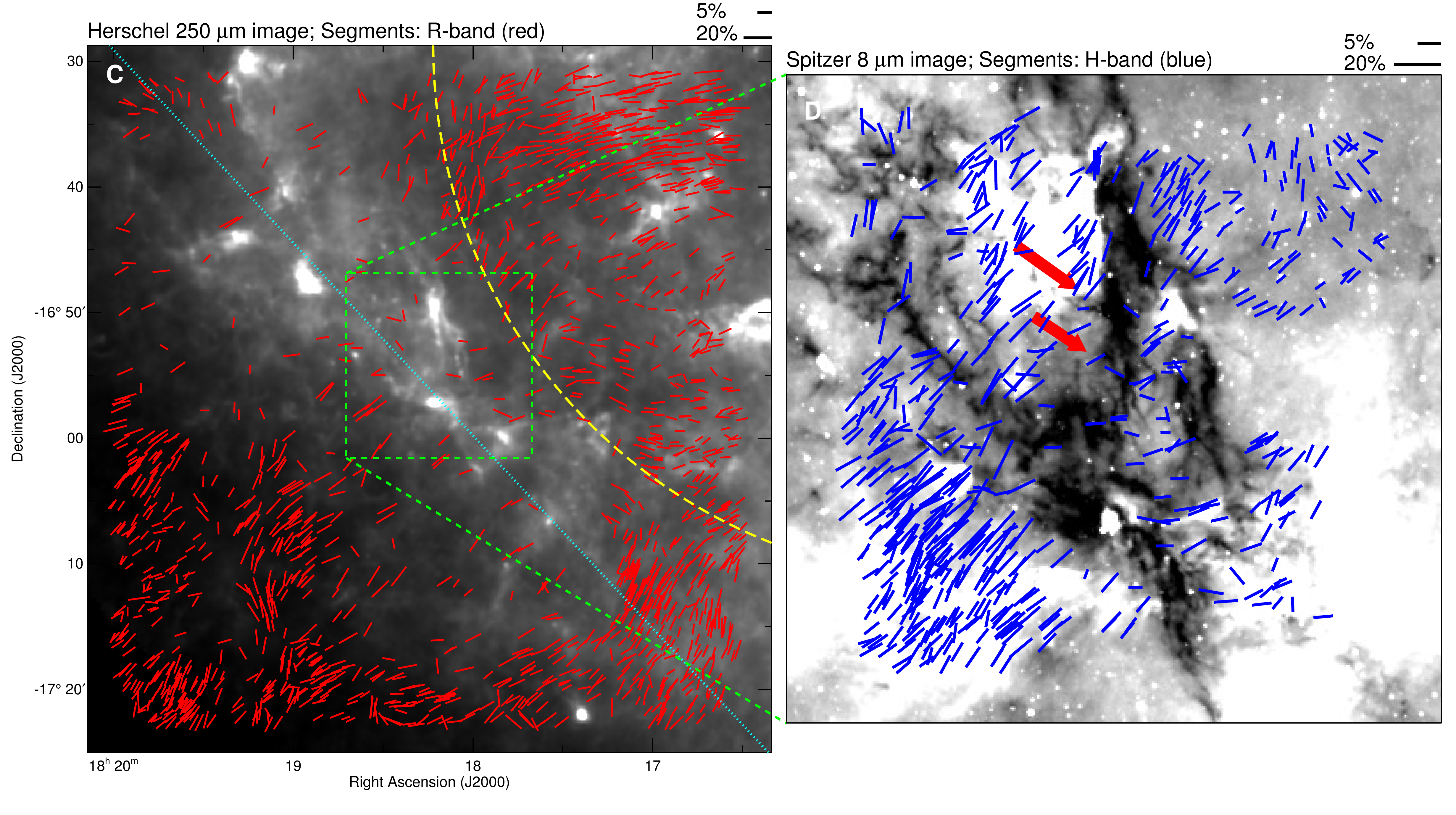} \\
      \caption{Polarization maps of IRDC G14.2 showing the entire R band ({\it a})
      and H band ({\it b}) datasets, as well as the foreground-corrected samples (see Section \ref{forecorrec})
      in the R and H band ({\it c} and {\it d}, respectively). The optical mapping (red segments) covers a large-area 
      around the cloud that is indicated by the cyan-colored $A_{V}$ contours in panel {\it a} 
      \citep[4, 5, 6 and 7 mag levels,][]{dobashi2013}. The NIR dataset (blue segments) is focused 
      on the central filamentary features (green contours in panel {\it b}, representing the integrated NH$_{3}$(1,1) emission 
      with levels of 40 mJy\,beam$^{-1}$\,km\,s$^{-1}$, from Paper I). 
      Polarization segments lengths are scaled proportional to $\sqrt{p}$ (each panel shows reference sizes 
      for $5\%$ and $20\%$ segments in the top right).
      Background images are from the SuperCOSMOS H$\alpha$ survey in panels {\it a} and {\it b} \citep{parker2005},
      from {\it Herschel}-SPIRE $250\,\mu$m in panel {\it c} and {\it Spitzer}-IRAC $8\,\mu$m in panel {\it d}. 
      The different wavelengths reveal emission features from different ISM components, 
      such as ionized gas, cold or warm dust. The RCW 157 area is indicated by the yellow dashed line in 
      panels {\it a} and {\it c}. Pillars at the edge of this area are located by the black arrow (see Section \ref{sec_rcw157}).
      Red arrows indicate striations perpendicular to the filaments, both in the H$\alpha$ ({\it b}) and 
      {\it Spitzer} images ({\it d}, see Section \ref{disc_filform}).
      The large-scale orientation of the cloud is indicated by the dotted cyan line in panel
      {\it c} ($\approx 43\degr$ relative to the north celestial pole). 
              }
         \label{polmap}
   \end{figure*}

Polarization orientations are assumed to trace the sky-projected orientation of magnetic 
field lines. To understand their relation to the surrounding ISM, we begin by plotting segments over different 
images covering distinct spectral ranges. 
Figures~\ref{polmap}a and \ref{polmap}b respectively show the entire ensembles of R-band (red) 
and H-band polarization data (blue). The segments sizes 
are proportional to $\sqrt{p}$, 
allowing a less biased visualization of the magnetic field morphology, 
particularly in this case where there is a mixture of segments displaying large variations 
of polarization degree. 

In this work, the analysis of the R-band and H-band polarimetric samples
are distinguished by the fact that they are useful in tracing respectively the large-scale
and the small-scale magnetic field structure around IRDC G14.2. More specifically, 
here we define small scales as the typical range of lengths of the 
filaments found in IRDC G14.2 ($\sim 1 - 3$ pc, green NH$_{3}$ contours in Figure~\ref{polmap}b), 
and large scales as sizes on the order of the 
molecular cloud in which the filaments are embedded ($\sim 20 - 40\,$pc, 
cyan visual extinction contours -- $A_{V}$ -- in Figure~\ref{polmap}a).
On one hand, while the R-band detections are limited by extinction to trace only more diffuse ISM,
they are distributed along a large area covering the molecular cloud's surroundings. 
On the other hand, the H-band polarimetry covers only the central areas, but are less affected 
by extinction and therefore a large number of segments are concentrated around the filaments.

In both Figures~\ref{polmap}a and \ref{polmap}b the background image corresponds to H$\alpha$ observations \citep{parker2005}. 
Thus, the image shows both stellar point sources and patches of bright extended emission due to the presence of the RCW 157 H{\sc ii} region, also known as Sh~2-44 (outlined by the curved yellow 
dashed line). The association of RCW 157 H{\sc ii} region with IRDC G14.2 is not clear as there is a discrepancy in the distance of RCW 157 region ($\sim2\,$kpc according to \citealt{avedisova1989}, and $3.7\,$kpc according to \citealt{deharveng2010} and \citealt{lockman1989}). In any case, in this work 
we excluded from the analysis the polarization data around RCW 157 region since the original morphology of magnetic field lines might 
have been distorted due to the expansion of the ionized volume. 
More discussion will be given in Section \ref{sec_rcw157}.

Figures \ref{polmap}c and \ref{polmap}d show the foreground-corrected polarization segments, respectively in the 
R and H bands. The detailed process of foreground correction is discussed in Section \ref{forecorrec}.
Background images in this case correspond to the {\it Herschel}-SPIRE $250\,\mu$m (Figure \ref{polmap}c) and 
the {\it Spitzer} $8\,\mu$m (Figure \ref{polmap}d). 
The large-scale dust cloud, as well as the complex of filamentary structures 
embedded within are clearly observed in these images. The close-up view from {\it Spitzer} (Figure \ref{polmap}d)
exhibits a better resolution view of the intricate pattern of interstellar filaments, seen in absorption against the 
Galactic background infrared radiation.

\subsection{Visual extinction estimates and foreground polarization correction}
\label{forecorrec}

   \begin{figure}[!ht]
   \centering
   \includegraphics[width=0.48\textwidth]{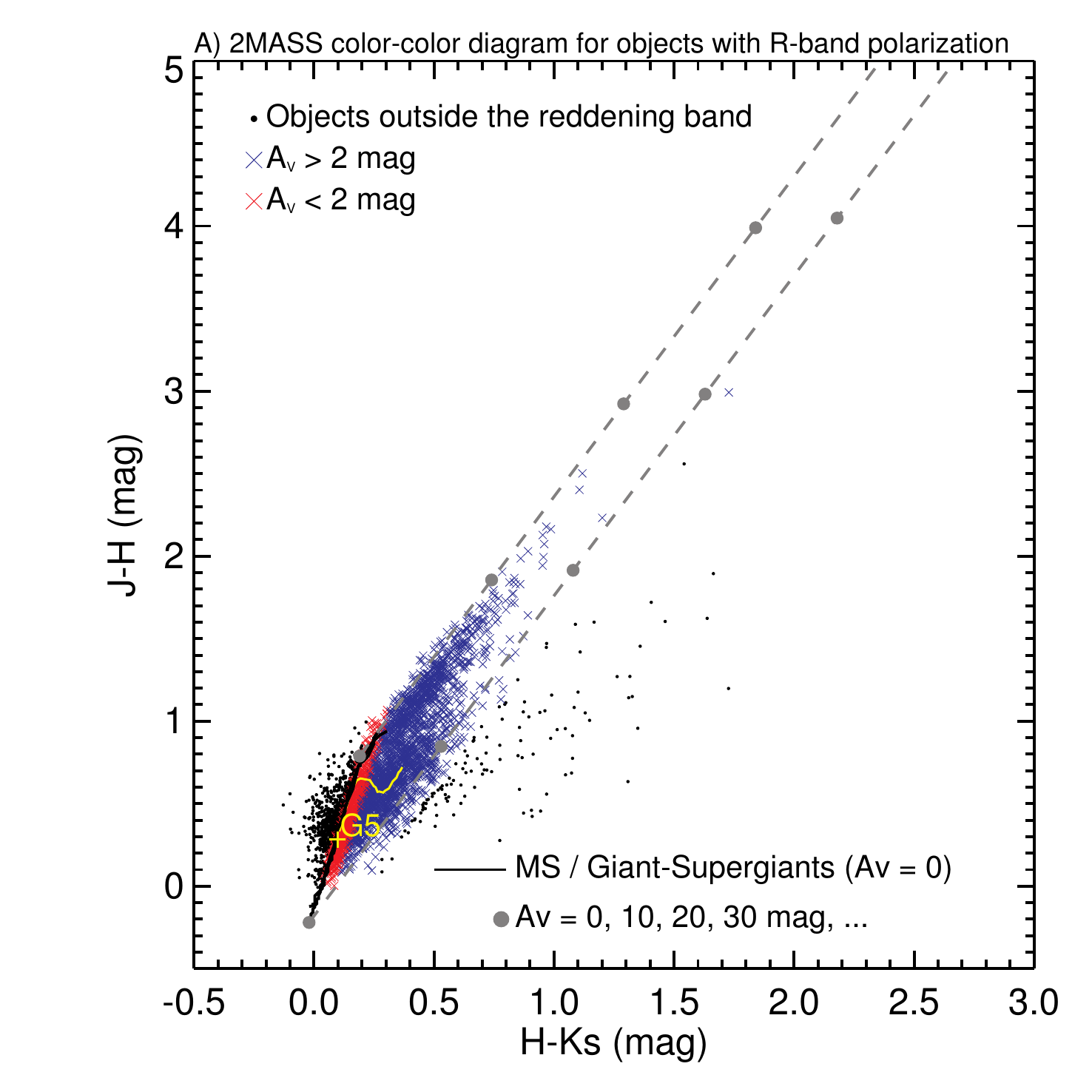} \\
   \includegraphics[width=0.48\textwidth]{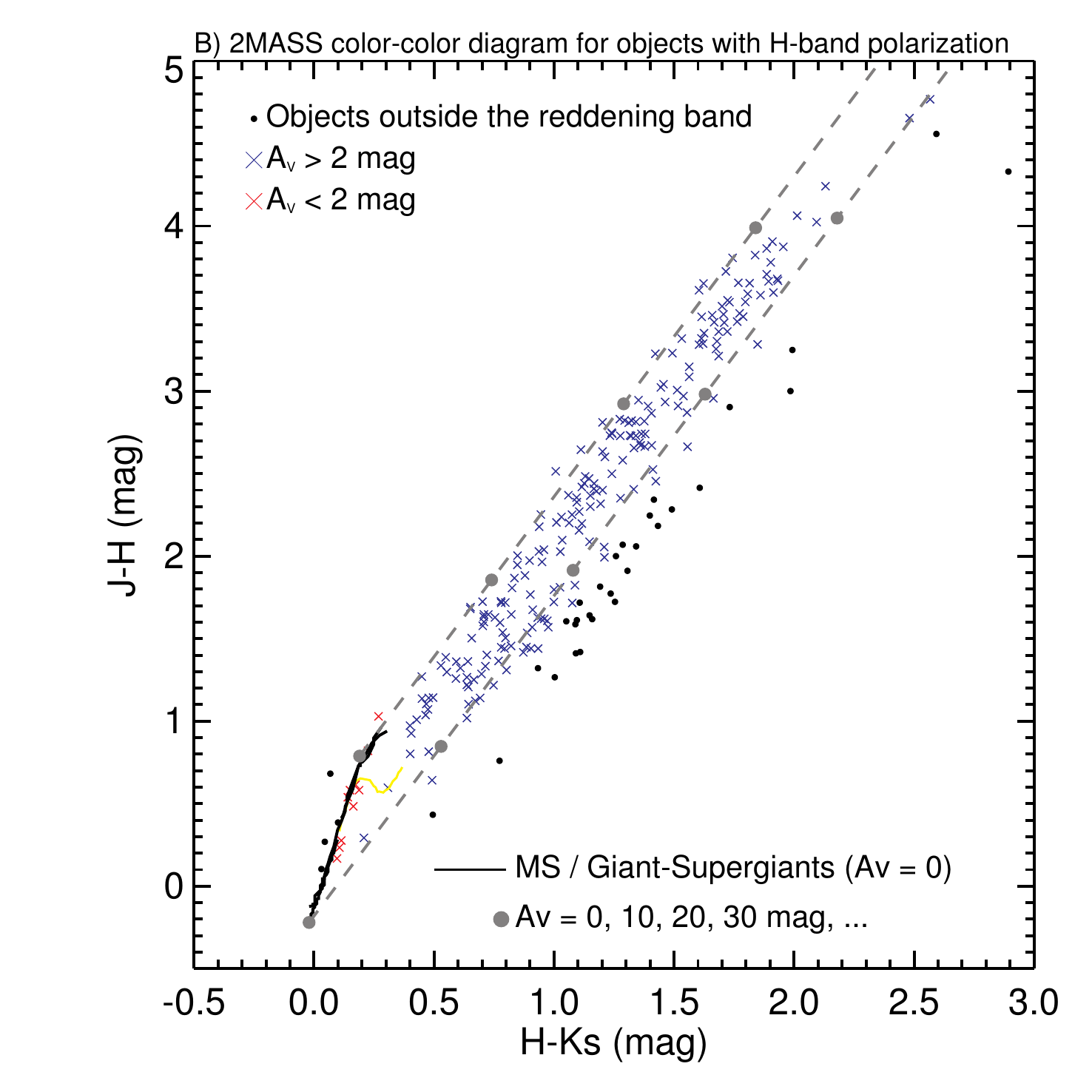} \\
      \caption{Color-color $(J-H) \times (H-Ks)$ diagrams based on 2MASS
               for objects from the R band ({\it a}) and H band datasets ({\it b}).
               The loci which corresponds to the unreddened ($A_{V} = 0$) main sequence, 
               giants and supergiants, is indicated by a solid line 
               \citep{koornneef1983,carpenter2001}. The reddening band \citep[parallel to 
               the reddening vector][]{fitzpatrick1999} is represented by dashed gray lines, in which
               each bullet indicated increments of 10 mag. The yellow line represents 
               main sequence stars with spectral types later than G5.
              }
         \label{ccdiag}
   \end{figure}

Considering the distance to IRDC G14.2, it is likely that a considerable fraction of the detections 
actually correspond to foreground stars (particularly those detected in the R-band mapping).
Therefore, two distinct operations must be applied to correct for the foreground contamination: 
\begin{itemize}
\item Correction A: foreground stars must be identified at least statistically, and removed from the sample; 
\item Correction B: the polarization component produced by the foreground material must be determined and 
subtracted from the background sources.
\end{itemize}
In the general direction of the dark cloud, 
stars distributed along different distances probe
interstellar polarization features produced by different interstellar components. 
Since individual distances are not known, one may use the visual extinction 
as a general proxy, giving us an approximate idea of the star's location along the line-of-sight.

Estimates of the visual extinction $A_{V}$ for each stellar object were obtained based 
on 2MASS photometric data \citep{skru2006}. Among the total of 4627 and 584 stars 
from our R-band and H-band samples, respectively, 1227 and 337 were either not found in 
the 2MASS catalog or excluded due to poor photometry in at least one of the NIR bands ($J$, $H$ or $Ks$). 
Thus, the following analysis applies only to the objects found in the 2MASS catalog, with
valid photometric values.

The visual extinction determination method is based on color-color diagrams
$(J-H) \times (H-Ks)$ which are shown in Figures \ref{ccdiag}a and \ref{ccdiag}b for the R-band and 
H-band polarimetric samples, respectively. As may be noted, reddening causes points
to spread along a band (gray dashed lines), since each data point is displaced from its de-reddened position
an amount proportional to the visual extinction. 
Therefore, by de-reddening each point upon reaching the main sequence locus, it is possible to 
estimate $A_{V}$ by applying general interstellar
relations given by \citet{fitzpatrick1999}. 
This method is not meant to provide a highly precise determination of $A_{V}$, since 
individual spectral types are not known and general assumptions regarding the relation between 
color excess and extinction have to be made \citep{fitzpatrick1999}. However, it 
is sufficiently robust to provide an approximate estimate, as needed in this work. 
It is important to point out that when de-reddening each point along the reddening band, 
the main sequence locus can 
be crossed twice (early-type and later-type stars), suggesting that there is an apparent 
degeneracy in the $A_{V}$ estimate. However, assuming the 2MASS photometric completeness limits, 
it is easy to show that un-reddened main sequence stars with spectral types later than G5 (the yellow line 
starting on the yellow plus sign) are too faint to be observed at such distances. Thus, the
late-type portion of the main sequence can be ignored (the yellow line) and only the early-type 
main sequence locus is used, removing the ambiguity. Also notice that the early-type portion 
of the main sequence locus is superposed to the giants and supergiants locus in a 
$(J-H) \times (H-Ks)$ diagram, and thus the $A_{V}$ estimate doesn't depend on the 
luminosity class.
For the R-band, only objects inside the reddening band are considered valid for this calculation (red or 
blue crosses), while objects outside (black dots) are excluded. In this way, sources with 
infrared color excess (typically displaced to the right side of the reddening band), 
which are known to present circumstellar discs (and therefore possibly intrinsic 
polarization by scattering), are automatically removed.


The R-band diagram shows that there are stars with a distribution of various extinction levels. 
Analyzing $E(b-y)$ reddening maps from \citet{reis2011} we notice that along the cloud's line-of-sight, 
the foreground ISM closer to the Sun
($d \lesssim 300$ pc) contributes with $A_{V} \sim 1$ mag (assuming the general relation $A_{V} = 4.3E(b-y)$). 
Estimates of the extinction and polarization levels
associated to the material beyond these local regions may be done by studying 
the compilation of $P_{V}$ (polarization degree at the $V$ band) and $E(B-V)$ (interstellar reddening) 
data by \citet{heiles2000} as a function of distance. 
Considering a radius of $1\degr$ centered on the cloud, 15 stars with distance smaller than $2.0$ kpc are found. 
Their mean R-band polarization degree and angle are respectively $0.7\pm0.2$\% and $67\degr$,
giving us an initial idea of the foreground polarization level.
It is important to point out that we have converted the polarization degree from V to R band using 
the relation by \citet{Serkowski1975}, assuming typical grain sizes, which corresponds to a 
peak in polarization spectrum around $\lambda_{max} = 0.55\mu$m.
The mean $E(B-V)$ value using the same 15 objects from \citet{heiles2000} is $0.6$ mag, corresponding to $A_{V} \approx 2.0$ mag 
(assuming $A_{V} = 3.1 E(B-V)$).

   \begin{figure}[!ht]
   \centering
   \includegraphics[width=0.48\textwidth]{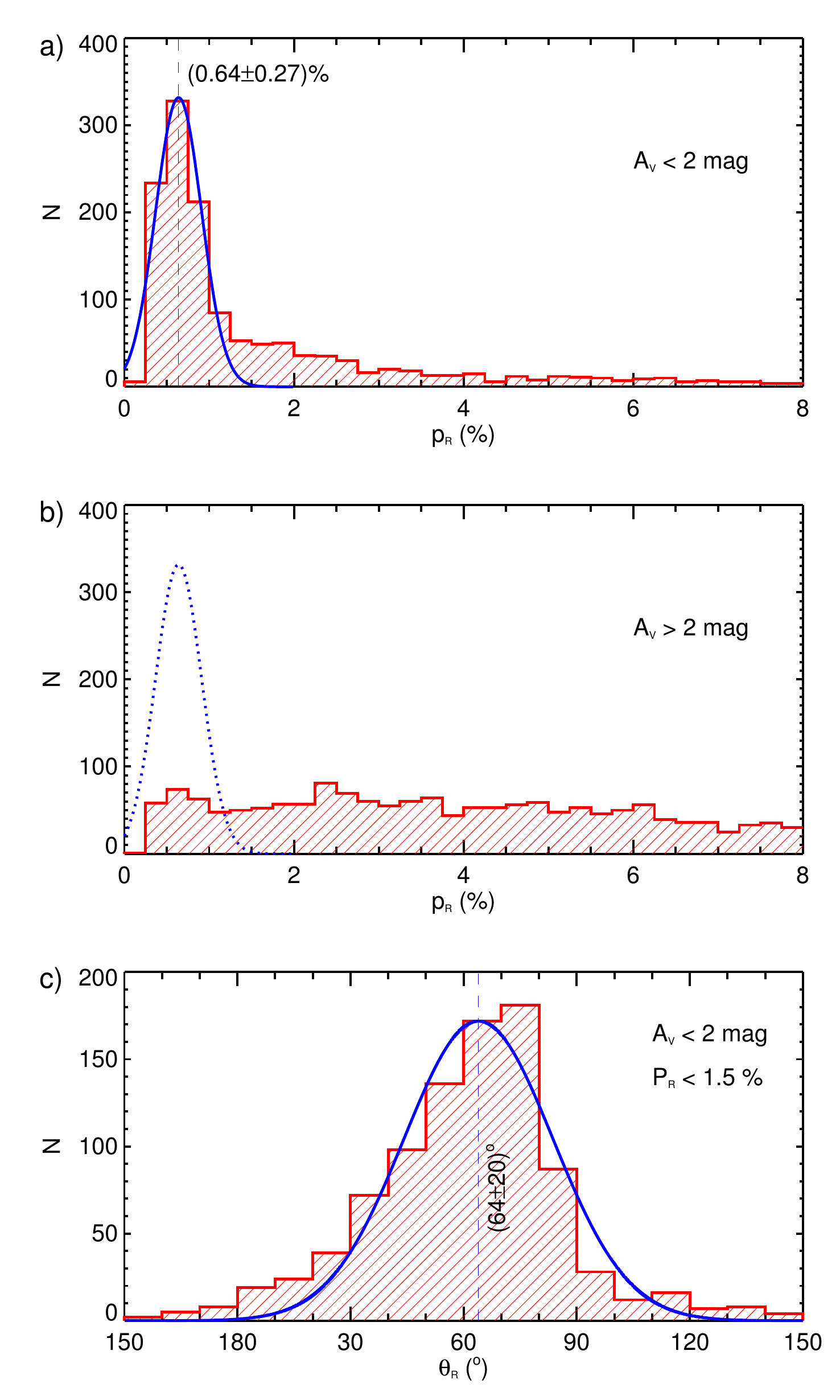} \\
      \caption{Histograms used to estimate the foreground contribution to the R band polarimetric
              sample: panel {\it a} shows the distribution of $p_{R}$ for $A_{V} < 2$ mag (which
              is a general proxy for the foreground extinction), with a Gaussian fit peaked at
              0.64$\%$; panel {\it b} shows the distribution for $A_{V} > 2$ mag 
              (the same Gaussian fit from panel {\it a} is shown as reference); 
              in panel {\it c} the distribution of $\theta_{R}$ is shown, considering only 
              polarization detections with $A_{V} < 2$ mag and $p_{R} < 1.5\%$. The Gaussian fit indicates an average 
              foreground polarization orientation of $\approx 64\degr$.
              }
         \label{fore}
   \end{figure}

Using the $2$ mag level as a general proxy for the foreground visual extinction, we proceed with the 
analysis by constructing the histograms in Figure \ref{fore}. The first histogram (Figure \ref{fore}a) 
shows the distribution of polarization values for stars with $A_{V} < 2$ mag, while Figure \ref{fore}b 
shows the distribution for $A_{V} > 2$ mag. A peak is seen in the first case (the blue Gaussian fit), while 
for higher extinctions (Figure \ref{fore}b) the Gaussian profile vanishes, shifting 
to a flat-like distribution. This indicates that objects encompassed by the 
Gaussian curve are probably foreground objects, while higher extinction sources are most likely background
stars. To determine the foreground polarization angle, the third histogram (Figure \ref{fore}c) shows
$\theta_{R}$ for stars with $A_{V} < 2$ mag and $p < 1.5\%$ (i.e., considering only objects below the 
Gaussian curve of Figure \ref{fore}a). From the peak of Gaussian fits of Figures \ref{fore}a and c, 
we estimate values of respectively 0.67\% and $64\degr$ for polarization percentage and orientation angle 
in the R band, matching very well the expectations based solely on the \citet{heiles2000} data. 
Additionally, the foreground value obtained is practically invariant under slight changes
in the $A_{V}$ and $p$ cut-offs used here, showing that this is a robust computation. 

Notice that the shape of the polarization angle distribution in Figure \ref{fore}c deviates slightly
from a Gaussian-like, suggesting that the foreground component is not perfectly uniform across the field.
This is not unexpected, given the wide field-of-view of the R-band survey area. The non-uniformity
probably corresponds to a smooth change in the foreground polarization angle across the field, 
since the distribution shows a unique non-symmetrical wide 
peak instead of multiple peaks clearly distinguishable.
Even though in this work we are adopting a single average foreground component, 
it is relevant to point out that for the purposes of the removal of this component from 
background sources (Correction B), the analysis that will be presented in Section \ref{s:largecomp} is very robust, 
and the same results are obtained even if no subtraction is applied (although Correction A is still
important).
The main reason is that the foreground component is usually small compared to polarization levels
of background stars, for which the molecular cloud component is predominant.

To apply Correction A, in order to be conservative in the selection 
of background sources, we consider only those with $A_{V} > 2$ mag
and $p > 2.0\%$ (i.e., those outside the range of the Gaussian fit from Figure \ref{fore}a), 
and we also exclude sources not found in 2MASS or rejected due to poor photometry.
For Correction B, we first calculate the mean foreground Q and U Stokes parameters using 
the mean foreground polarization that was previously obtained ($p_{V}=0.67\%$ and $\theta_{V}=64\degr$).
Then, we subtract this mean foreground Q and U value from each 
background star, finally determining a sample 
of foreground-corrected R-band detections which are probably mostly composed of background sources.
The polarization segments for the foreground-corrected sample is shown in Figure \ref{polmap}c. 

In the case of the H-band sample, the color-color
diagram (Figure \ref{ccdiag}b) shows that only a few stars are low extinction sources 
($A_{V} < 2$ mag). These few objects are excluded from the final sample, lending the map from 
Figure \ref{polmap}d, in which most sources are probably from the background, given 
their $A_{V}$ levels. The small fraction of foreground stars found in the H-band
dataset with 2MASS data suggests that even considering the entire dataset (including 
objects not found in 2MASS or excluded due to poor photometry), the vast majority of 
stars are probably background sources. Thus, we consider objects not found 
in 2MASS (or rejected) as background sources for the H-band polarization analysis in this work. 
For subtraction of the foreground component from background sources (Correction B), 
we find that in the H band the contribution is negligible: if $p=0.67\%$ in the R band, 
then assuming the Serkowski relation, the H-band foreground polarization would be approximately
0.15\%. Since this is a small level of polarization, lower than the typical uncertainty  
in polarization degree, we choose to ignore its contribution. This avoids introducing unnecessary
systematic uncertainties, since the estimate of $0.15\%$ for the 
H-band foreground polarization (extrapolating from the R band) involves assumptions regarding the
peak of the polarization spectral function.

   \begin{figure}
   \centering
   \includegraphics[width=0.48\textwidth]{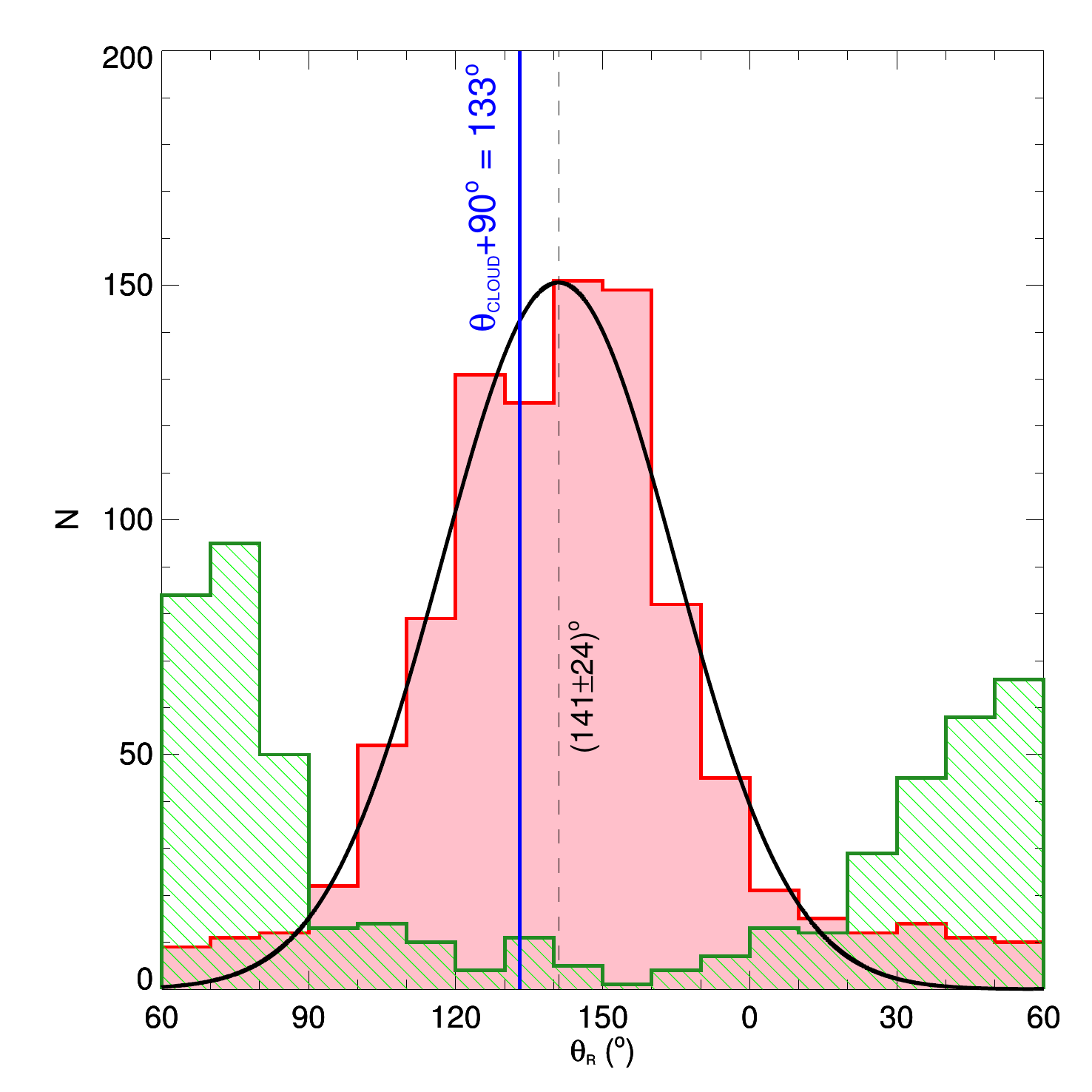} \\
      \caption{Histogram (red) of the foreground-corrected orientation of polarization
               segments in the R-band (as shown in Figure \ref{polmap}c), excluding 
               the RCW 157 region. The peak of the distribution ($\approx 141\degr$) is found through 
               the Gaussian fit, being approximately perpendicular to the large-scale 
               cloud orientation ($\theta_{\mathrm{cloud}} + 90\degr = 133\degr$).
               The green histogram correspond to foreground objects removed from 
               the sample (see Correction A in Section \ref{forecorrec}).
              }
         \label{compcloud}
   \end{figure}

\subsection{Relation between polarization segments and the large-scale cloud orientation}
\label{s:largecomp}

After removal of the foreground stars from the R-band sample, 
it is possible to investigate the relation between the orientation of 
polarization segments and the large-scale cloud in which the interstellar filaments are embedded. 
This may help determine the range of spatial scales in which magnetic fields might be 
important in regulating the gravitational collapse. 

Figure~\ref{compcloud} shows a histogram of R-band polarization angles (red), excluding 
the area defined by the RCW 157 H{\sc ii} region (above the dashed yellow line in 
Figure~\ref{polmap}a). Although there is a large dispersion, a peak around $141\degr$ 
is clearly identified (as shown by the Gaussian fit). In comparison, the direction 
perpendicular to the cloud ($\theta_{\mathrm{cloud}}+90\degr$) is indicated 
by the blue line, as $\sim133\degr$ (the cloud's direction, $\approx43\degr$, is shown by the cyan-colored 
dotted line in Figure~\ref{polmap}c). It is clear that there is an overall correlation 
between the large-scale magnetic field lines and the direction perpendicular to the 
cloud.

It is important to point out that, particularly 
for this analysis, the previous removal of foreground sources 
was essential (Correction A), since these comprised a considerable fraction of the vector sample. 
The green histogram in Figure \ref{compcloud} shows foreground stars that were removed 
from the sample. Notice that foreground segments
in general are parallel to the cloud, which is an opposite trend compared to 
background sources. 
Comparing Figures \ref{polmap}a and c, it is straightforward to visualize the 
sample of foreground stars that has been removed (which are mostly low polarization detections
parallel to the cloud orientation).
Therefore, if not previously removed, this component 
would have introduced considerable contamination in this analysis, impairing the notion 
that on-site magnetic field lines in general are perpendicular to the large-scale cloud.

\subsection{Relation between polarization segments and the orientation of filaments}

\begin{figure*}[!ht]
   \centering
   \includegraphics[width=0.48\textwidth]{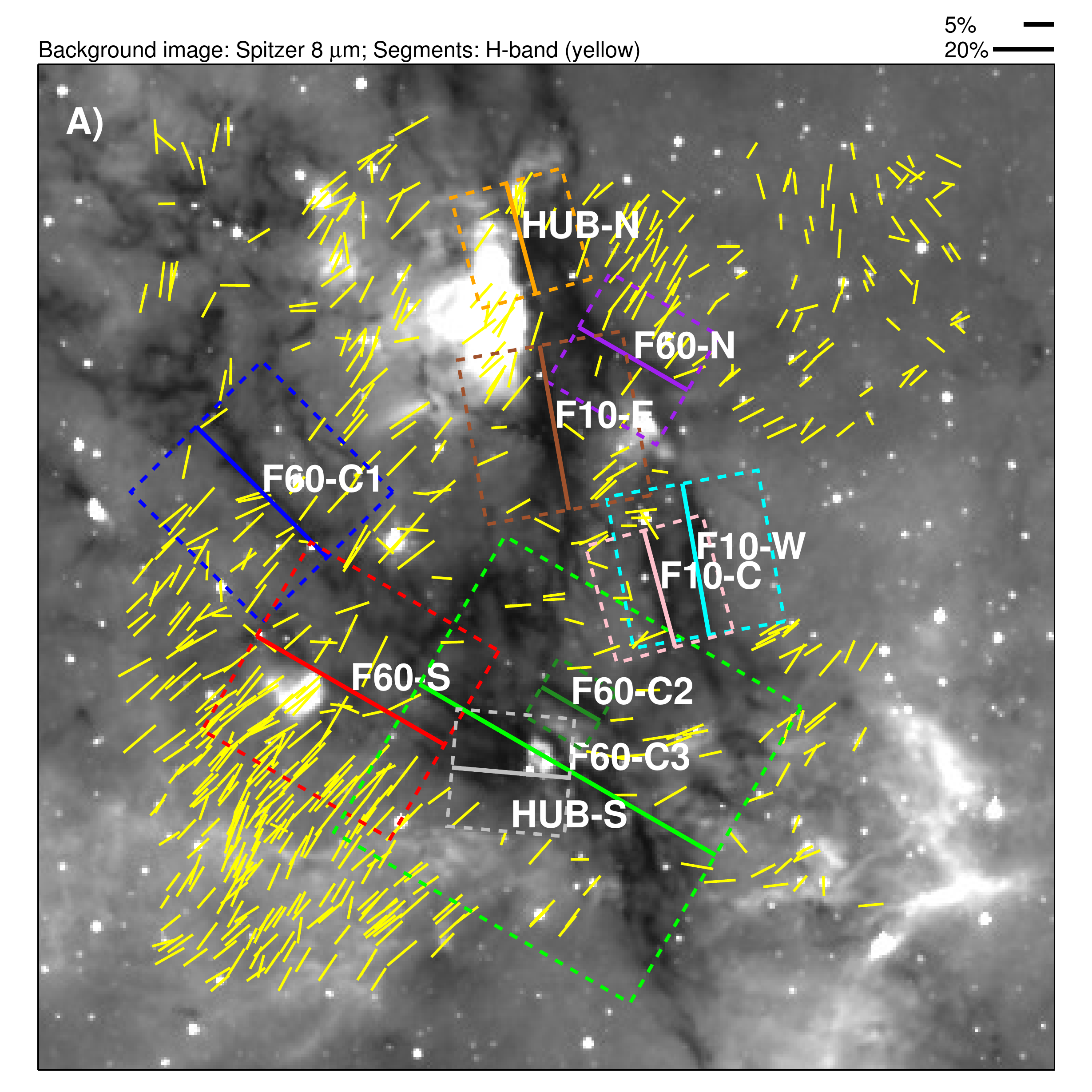} 
   \includegraphics[width=0.48\textwidth]{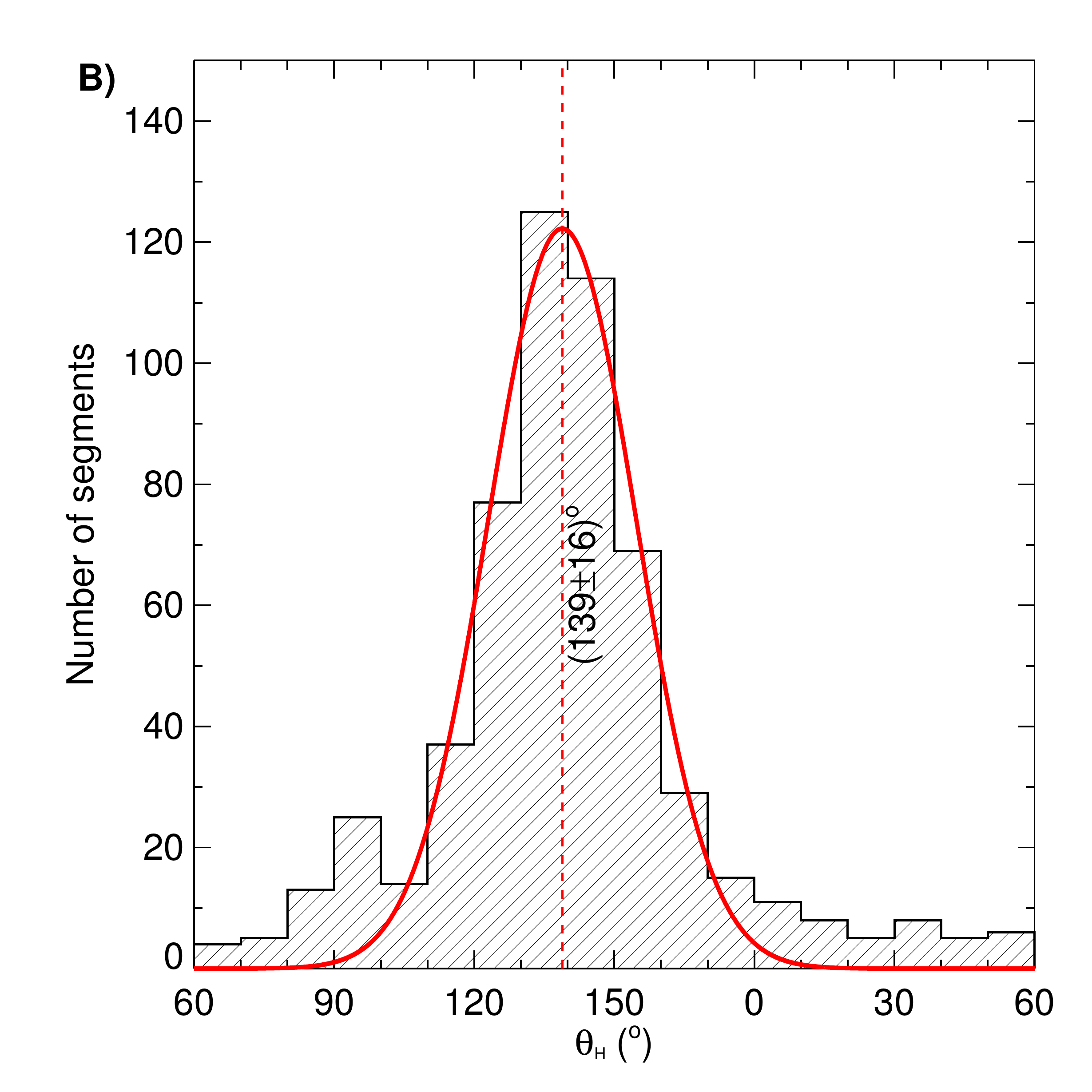} \\
   \includegraphics[width=0.48\textwidth]{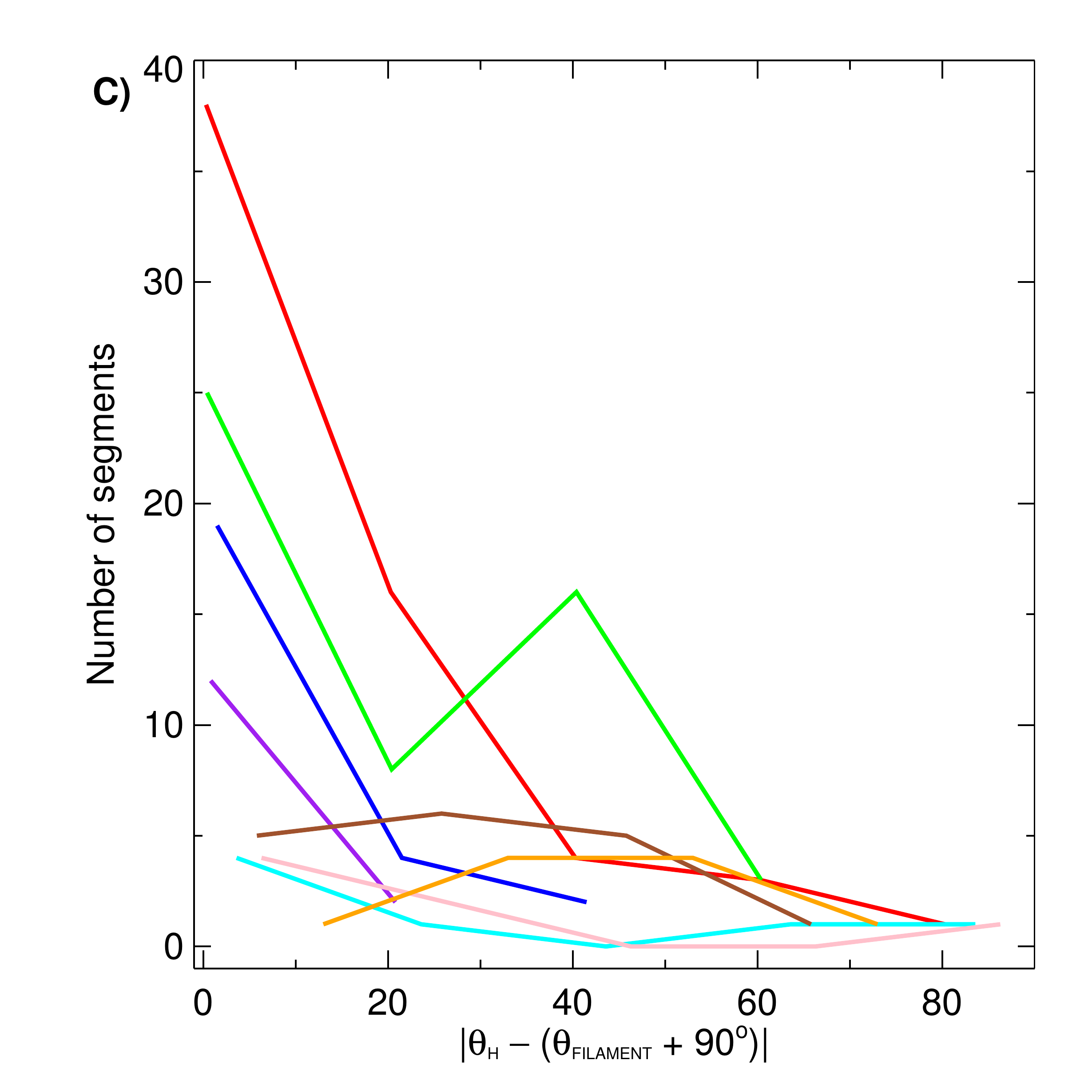}  
   \includegraphics[width=0.48\textwidth]{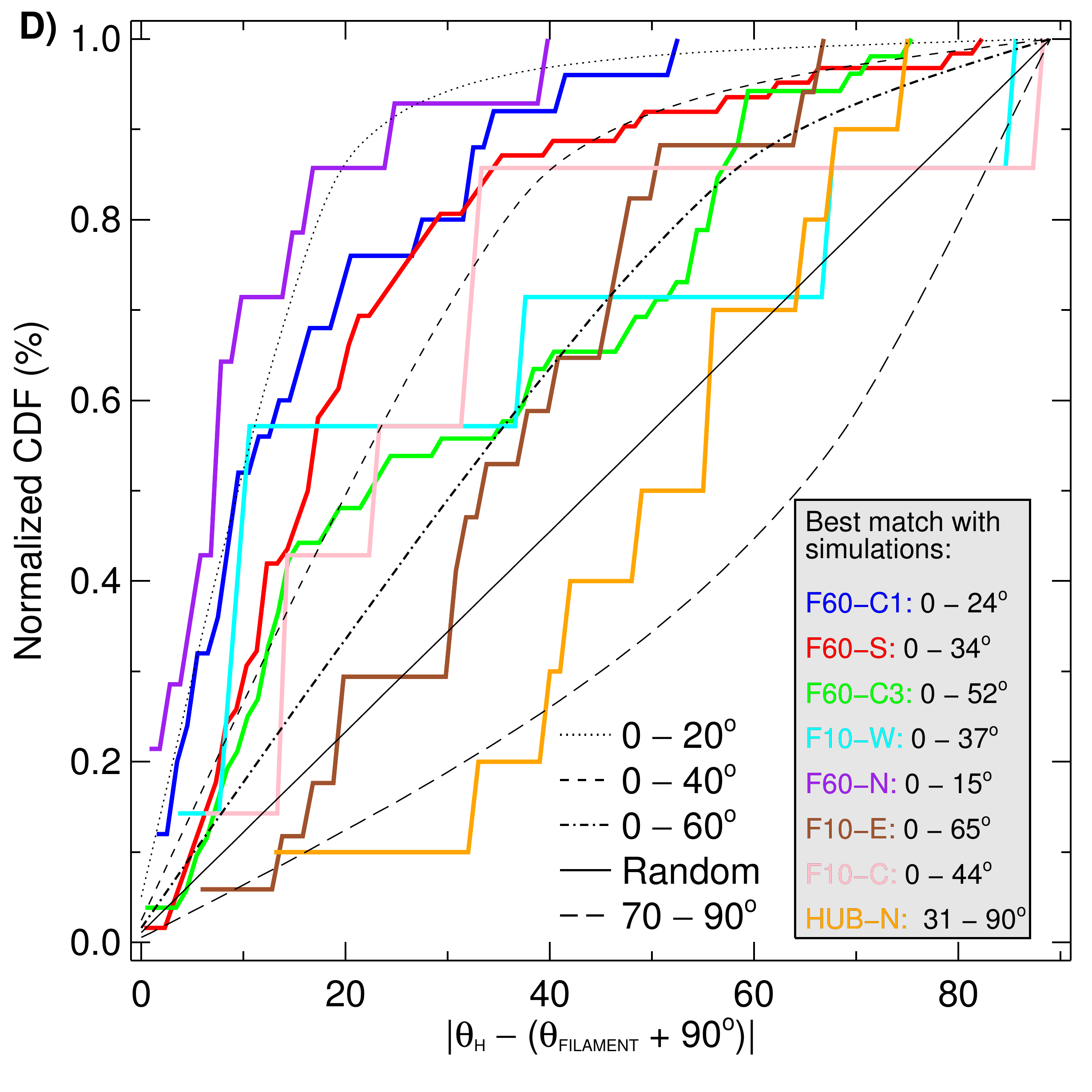} \\
      \caption{Analysis of the relative orientation between H band 
              polarization segments and interstellar filaments. 
              Panel {\it a} shows the {\it Spitzer} $8\mu$m image with the 
              polarimetric data (same as Figure \ref{polmap}d), as well as dashed colored boxes with sizes equal to each 
              filament's length (represented by the solid lines, 
              as defined in Paper I).
              Panel {\it b} shows the histogram of polarization orientation for 
              all H-band detections in the field.
              Panels {\it c} and {\it d} respectively show the regular histograms and
              cumulative distribution functions (CDFs) of the difference 
              between polarization angle and the orientation perpendicular to each filament, 
              considering only the segments within the boxes (the colors match each box in panel {\it a}).
              Black lines shows numerical predictions for the CDFs, as based on Monte Carlo simulations.
              Preferential ranges for tridimensional angle differences of $\Delta a = 0-20\degr$, $0-40\degr$,
              $0-60\degr$ and $70-90\degr$ are indicated as reference, as well as a completely random distribution
              (the solid line). The best $\Delta a$ match with the simulations for each filament are listed in the gray box.
              }   
         \label{compfilaments}
   \end{figure*}

Figure \ref{compfilaments}a shows the H-band polarization segments superposed to the 
{\it Spitzer} $8\,\mu$m image, together with the location of filaments represented by colored straight lines.
In Paper I, these structures were distinguished between hubs and filaments depending 
on physical features obtained from the NH$_3$ observations:
hubs were classified as structures presenting signs of star formation, as well as higher rotational temperatures and non-thermal 
velocity dispersions ($T_{\mathrm{rot}} \sim 15\,$K and $\sigma_{\mathrm{NT}} \sim 1\,$km s$^{-1}$) as compared to filaments 
($T_{\mathrm{rot}} \sim 11\,$K and $\sigma_{\mathrm{NT}} \sim 0.6\,$km s$^{-1}$). 

Figure \ref{compfilaments}b shows a histogram of polarization orientation in the H band ($\theta_{H}$), which includes all 
the detections shown in Figure \ref{compfilaments}a. It clearly exhibits
a peak at $\theta_{H} = 139\degr$. It is interesting to note that the main orientation at such smaller
scales matches very well the average orientation at large-scale (from the Figure \ref{compcloud}).

The relative orientations of segments and filaments are 
projected into the plane of the sky, so the true relative orientations are unknown.
To carry out a quantitative analysis of the relative orientations, a box was drawn 
around each filament, with sizes matching the length of each structure (from Paper I). Thereafter, segments inside each 
box were selected in order to represent the orientation of magnetic field lines in the 
immediate surroundings of each filament. For each vector, its orientation relative 
to its corresponding filament's perpendicular direction was computed ($|\theta_{H} - (\theta_{\mathrm{filament}}+90\degr)|$).

Figures \ref{compfilaments}c and \ref{compfilaments}d respectively show the regular distributions
and the Cumulative Distribution Functions (CDFs), using the relative orientations of
segments for each filament (Hub-S and F60-C2 were omitted).
The histograms and CDFs are shown with colored lines that match each respective filament (and its box).
Although there is considerable variation in the orientation of polarization segments throughout the 
field, there is a clear trend for an overall orientation perpendicular to the filaments.
This can be seen by the peaks close to zero for some of the histograms in Figure \ref{compfilaments}c.

In order to account for the possible geometrical projection effects, we compared the CDFs to Monte Carlo 
simulations of a set of relative projected angles based on a large number of vector-filament pairs 
randomly distributed in tridimensional space. For each individual simulation, we selected 
only pairs in which the true relative orientation was within a certain range of values 
(denoted by $\Delta a$).
Using this subset of segments, we projected the pairs in the plane of the sky and then computed 
the CDF of the projected relative orientations. 
Examples for $\Delta a$ equal to $0 - 20\degr$, $0 - 40\degr$, $0 - 60\degr$, and 
$70 - 90\degr$ are shown in Figure \ref{compfilaments}d,
as well as the random condition (or $0 - 90\degr$).

To find out which $\Delta a$ configuration from the simulations 
would best represent the segments' orientations for each filament, 
we begin by running it for all possible $\Delta a$ ranges.
Then, for each filament, we compare its observed CDF to each of the various simulated CDFs
through Kolmogorov-Smirnov tests, which are useful to verify the statistical probability of two 
different distributions be drawn from the same ensemble. Finally, 
the comparison which provided the larger probability was chosen as 
the best simulation that could represent the observed CDF. The $\Delta a$ for
the best representative simulation for each filament is shown in 
Figure \ref{compfilaments}d.

As expected by visual inspection, the majority of the filaments present $\Delta a$
upper limits significantly lower than $90\degr$. This means that there is a very clear trend of 
filaments and hubs being perpendicular to magnetic field lines, even when considering 
that both the filaments and the polarization segments orientations represent 
a projection in the plane of the sky. There are, however, some 
situations where the statistics is not ideal (for example, the small number of detections
for F10-C and F10-W) and a few exceptions, 
for example: for Hub-N, the best representative simulation
corresponds to a $\Delta a$ range between $31$ and $90\degr$, suggesting a slight 
trend of magnetic field lines parallel to the hub. In addition, the distribution
for F10-E is only marginally representative of a perpendicular condition. It is 
interesting to notice that these discrepancies occur exactly for the two structures
that are spatially closer to IRAS$\,$18153-1651, the bright ultra-compact H{\sc ii} region
to the east of Hub-N and F10-E. This suggests that magnetic field lines in these 
structures might have been disrupted by the H{\sc ii} region expansion.
Further discussion is given in Section \ref{disc_filform}.

\subsection{Statistical derivation of the magnetic field strength}

In order to understand the interplay between magnetic field support, gravity and turbulence 
for each filamentary structure, important physical parameters may be calculated 
by combining the H-band polarization data with velocity dispersion data from molecular-line 
studies 
and density information. 
These parameters are the plane-of-the-sky component of the magnetic field strength ($B_{\mathrm{pos}}$), the Alfv\'en Mach number 
($M_{A}$) and the mass-to-magnetic-flux ratio ($\lambda$). 

Given a set of polarization segments surrounding a certain filament or hub, the 
Chandrasekhar-Fermi (CF) theory \citep{chandrasekhar1953} states that the magnetic field strength 
in that volume of the ISM is inversely proportional to the angular dispersion of polarization 
segments, a quantity that is related to turbulence. 
A quantitative method may be applied to study such angular dispersion factor, 
which represents the signature of interstellar turbulent motion impinged in the morphology of 
magnetic field lines in that area. 
The method consists in a statistical analysis,
proposed first by \citet{hildebrand2009} and extended later on by \citet{houde2009}, which takes into account the effect of the line-of-sight depolarization. This method has been successfully applied to optical polarization data \citep{franco2010} as well as to submillimeter polarization data \citep[e.g.,][]{houde2011,girart2013,frau2014}. 

As shown by \citet{houde2009} the angular dispersion function (ADF) can be used to estimate the importance of the magnetic field. 
We have estimated the angular dispersion function, $1-\langle\mathrm{cos}[\Delta\Phi(l)]\rangle$, where $\Delta\Phi(l)$ is the difference in polarization angles between two points in the plane of the sky separated by a distance $l$. The analysis is based on the assumption of a stationary, homogenous, and isotropic magnetic field strength and a magnetic field turbulent correlation length, $\delta$, smaller than the thickness of the cloud $\Delta'$. Under these assumptions, the angular dispersion function (Equation~(42) from \citealt{houde2009}) can be expressed as

\begin{eqnarray}
1-\langle\,\mathrm{cos}[\Delta\Phi(l)]\rangle&\simeq&\frac{\langle\,B_{\mathrm{t}}^2\rangle}{\langle\,B_{\mathrm{0}}^2\rangle}\frac{1}{N}[1-e^{l^2/2(\delta^2+2W^2)}] \nonumber \\
&&+ \sum_{j=1}^\infty\,a_{2j}^{'}l^{2j},
\label{e:adf}
\end{eqnarray}

\noindent where $l$ is the length scale, $W$ is the standard deviation of the Gaussian beam ($W=FWHM/\sqrt{8\,\mathrm{ln}\,2}$), $\delta$ is the turbulent correlation length, and $N$  is the number of independent turbulent cells along the line of sight, 

\begin{eqnarray}
N=\biggl[\frac{(\delta^2+2W^2)\Delta'}{\sqrt{2\pi}\delta^3}\biggr].
\label{e:cells}
\end{eqnarray}

\noindent The summation term represents the contribution from the ordered component of the magnetic field that does not 
involve turbulence. 
The coefficient $a_{2j}$ represents to the steepness of the function in this ordered component. 
For stellar polarimetry data, the beam size can be considered as a pencil beam, since $W$ is negligible relative 
to the turbulent length scale $\delta$ (thus $W$ may be ignored).
The intercept of the fit to the data of the uncorrelated part at $l=0$, $f_{\mathrm{NC}}(0)$, allows us to estimate the turbulent to large-scale magnetic field energy ratio ($\langle\,B_{\mathrm{t}}^2\rangle/\langle\,B_{\mathrm{0}}^2\rangle$) as

\begin{eqnarray}
\frac{\langle\,B_{\mathrm{t}}^2\rangle}{\langle\,B_{\mathrm{0}}^2\rangle}=N&f_{\mathrm{NC}}(0).
\label{e:fnc}
\end{eqnarray}


   \begin{figure}[t]
   \centering
   \includegraphics[width=0.45\textwidth]{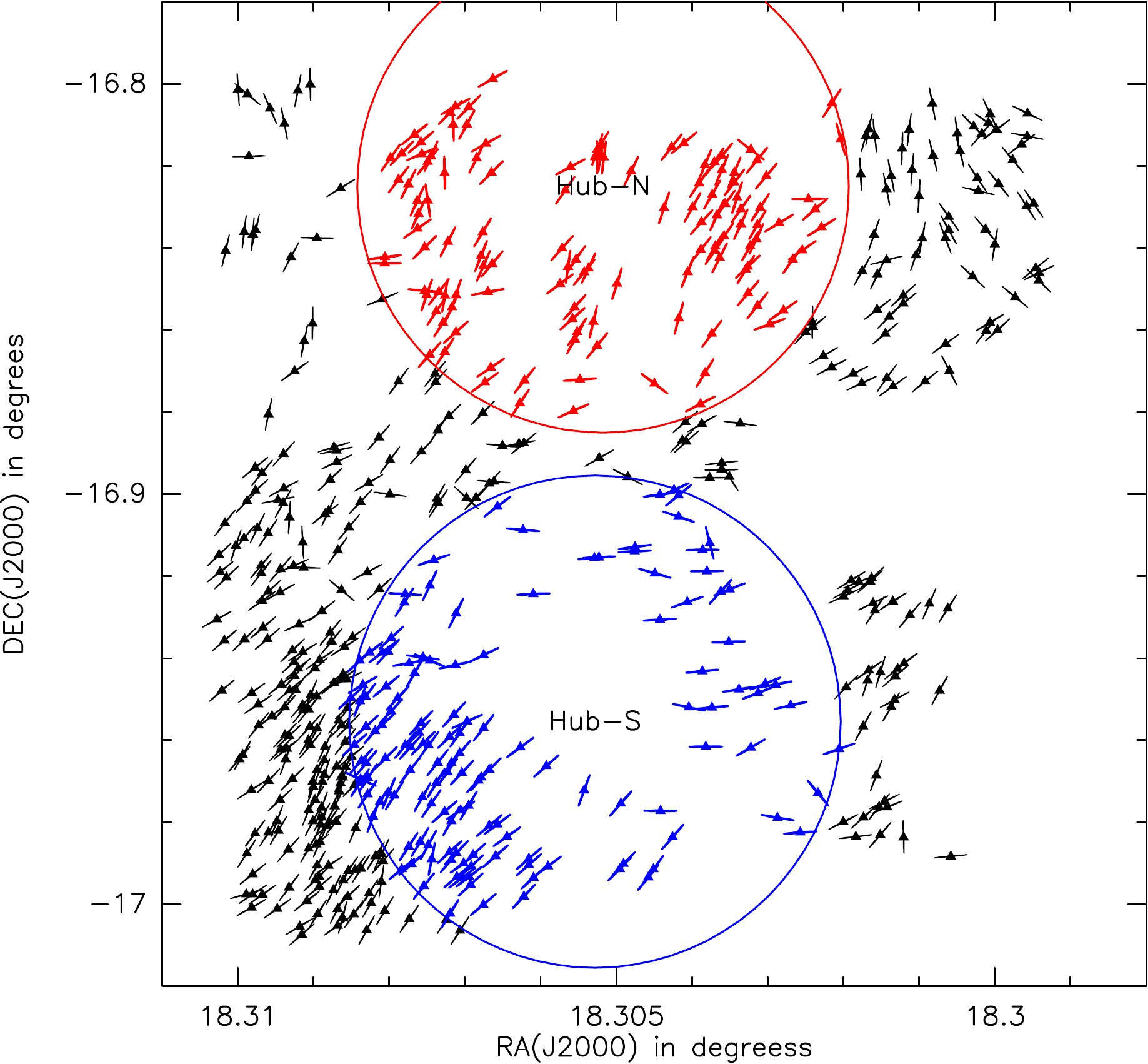} \\
\caption{Magnetic field direction obtained from H-band polarization data in IRDC G14.2. Red and blue segments indicate the polarization data used to compute the angular dispersion function around Hub-N and Hub-S, respectively. }
\label{pol-regions}
   \end{figure}
   \begin{figure}[t]
   \centering
   \includegraphics[width=0.45\textwidth]{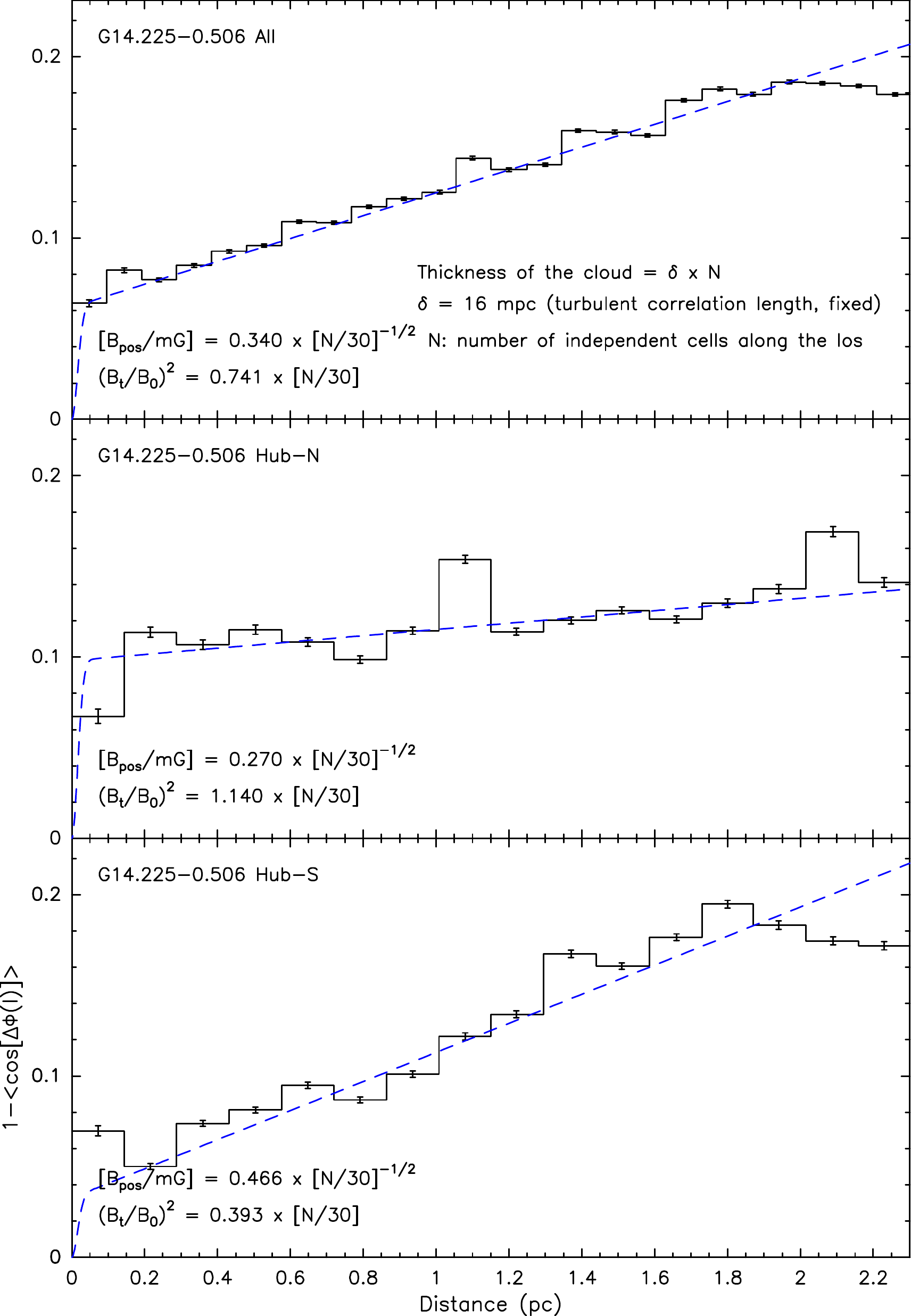} \\
\caption{Angular dispersion function of the magnetic field segments detected toward the IRDC G14.2 considering all B-field segments (top), Hub-N (middle), and Hub-S (bottom) considering, respectively, the red and blue B-field segments shown in Fig.~\ref{pol-regions}. The data points and error bars are the mean and standard deviation of all the pairs contained in each bin. The blue dashed line shows the best fit to the data (Equation~\ref{e:adf}).}
\label{adf}
   \end{figure}

   The low statistics obtained in the IRDC G14.2 prevent us to conduct the statistical analysis to fit the ADF for each filament and hub individually. Instead, to analyze the magnetic field, we considered three different regions: all the cloud, and the two hub-filament systems identified in Paper~I, Hub-N and Hub-S. We defined a radius, $R=0.06\degr$ or $\sim2$~pc, from the center of each hub \citep{busquet2013,busquet2016} to estimate the angular dispersion function for all the measurements that are at a distance $<R$ of the hub. Figure~\ref{pol-regions} shows the circles centered in each hub for this radius, indicating the polarization values used to compute the ADF for Hub-N (in red) and Hub-S (in blue). The radius of $0.06\degr$ was chosen using the following criteria: (1) to make sure sufficiently wide areas around Hubs N and S were covered, while also avoiding an overlap between them; (2) to avoid including in the Hub-N area a group of polarization segments to the right of the red circle that clearly show a different mean orientation, probably related to the edge of RCW 157 (compare with Figures \ref{polmap}a and b).
In Figure~\ref{adf} we present the angular dispersion function for all the cloud (top panel), Hub-N (middle panel), and Hub-S (bottom panel). One may notice that each function consists of a gradual rise starting from $l=0$, which may be interpreted as a decrease in the correlation of polarization orientation for segments separated by increasingly larger angular distances. 
The behavior of the ADF is slightly different in the two regions defined around each hub, with Hub-N having a more flattened slope than Hub-S, indicating that the large-scale magnetic field in the plane of the sky is quite uniform. 
The best fit of Eq.~\ref{e:adf} to the polarimetric data is shown in Fig.~\ref{adf} with the blue dashed line. 

To calculate $\langle\,B_{\mathrm{t}}^2\rangle/\langle\,B_{\mathrm{0}}^2\rangle$, we begin by estimating $N$, which is related
to the cloud thickness along the line of sight $\Delta'=\sqrt{2\pi}\,N\delta$ (see equation \ref{e:cells}). 
In other star forming regions, the turbulent correlation length $\delta$ was found to be equal to $16$~mpc 
(OMC-1, \citealt{houde2009}; DR21-OH, \citealt{girart2013}), or varying between $13$ and $33$~mpc in 
NGC7538 IRS1 \citep{frau2014}. Based on these previous estimates, in this work we fix $\delta = 16$~mpc 
since it is not the main source of uncertainty, as will be noted below.

The cloud thickness can be estimated by taking the ratio between the column density $N_{\mathrm{H_2}}$
and the volume density $n(\mathrm{H_2})$. We should point out that both quantities are estimated here
for the material surrounding the filaments, to coincide with the region where the H-band 
polarization data is distributed. The volume density is the main source of uncertainty 
for this calculation, so the approach is to find reasonable lower and upper limits around the filaments, 
and use this range as a proxy to determine the uncertainty in the magnetic field strength.
For the lower limit, we notice that the C$^{18}$O\,(1-0) line data from IRAM 30m (Busquet et al. in prep.) 
reveals an emission present over the entire IRDC G14.2 field, covering not only the dense filaments but 
also their surroundings. Thus, a conservative estimate for the lower limit is the critical density
of C$^{18}$O\,(1-0) which is $\sim1.4\times10^{3}$~cm$^{-3}$ \citep{1999myers}. From the same molecular line survey, we 
find that the HCN\,(1--0) line is also detected in the more diffuse area between filaments, 
thus its effective excitation density $4.5\times10^{3}$~cm$^{-3}$ (assuming a temperature of $20$~K, see Table~1 
of \citealt{shirley2015}), is representative of the typical density in this material. For the 
upper limit, we know that the density cannot be too much higher than $10^{4}$~cm$^{-3}$, because 
molecular line transitions with higher excitation densities (such as the HC$_3$N\,(10--9) line, 
with and excitation density of $4.3\times10^{4}$~cm$^{-3}$ at $20$~K) are found in emission only 
toward the densest portions of the filaments. Therefore, we adopt the range of $n(\mathrm{H_2})$
between $\sim1.4\times10^{3}$~cm$^{-3}$ and $10^{4}$~cm$^{-3}$, and propagate the uncertainties 
into the cloud thickness, $\langle\,B_{\mathrm{t}}^2\rangle/\langle\,B_{\mathrm{0}}^2\rangle$, 
and the magnetic field strength.

The column density ($N_{\mathrm{H_2}}$) can be estimated for the inter-filament region 
by two independent methods: 
(1) Using multi­wavelength dust emission maps (from ground-based -- CSO, APEX -- and space telescopes -- 
Herschel, Planck) to carry out a single component, modified black-body fit to each pixel of the maps
(Lin et al. 2016, in prep.). The derived values for the region sampled by NIR polarization 
are typically $T\approx20$~K and $N_{\mathrm{H_2}}\approx10^{22}$~cm$^{-2}$;
(2) Using the RADEX\footnote{http://var.sron.nl/radex/radex.php} on-line one-dimensional 
non-LTE radiative transfer code \citep{2007vandertak} to obtain the 
column density based on the C$^{18}$O\,(1-0) line. As inputs to the line data model, we used a line
width of $\sim2$~km\,s$^{-1}$, temperatures of $20$~K and volume densities in the 
range of $\sim1.4\times10^{3}$~cm$^{-3}$ to $10^{4}$~cm$^{-3}$. These inputs result in C$^{18}$O
column densities between $2\times10^{22}$~cm$^{-2}$ and $5\times10^{22}$~cm$^{-2}$. 
Assuming the standard $^{16}$O/$^{18}$O ratio for the local ISM of 560 \citep{wilson1994}, 
and adopting the standard abundance of CO with respect to H$_{2}$ of $10^{-4}$, we find $H_{2}$
column densities in the range $1 - 2 \times 10^{22}$~cm$^{-2}$.
Therefore, $N_{\mathrm{H_2}}$ is well constrained by two independent methods to be $\approx10^{22}$~cm$^{-2}$, 
and we adopt this as a fixed value to obtain the cloud thickness. The range of cloud thickness
is between $0.32$ and $2.31$~pc, yielding a number of independent cells ranging from $N = 10$ to $60$.
Using an average value of $N = 35$, this implies that $\langle\,B_{\mathrm{t}}^2\rangle/\langle\,B_{\mathrm{0}}^2\rangle$ 
in all the cloud is 0.86, while 
the values are 1.33 and 0.46 in Hub-N and Hub-S, respectively (see Table~\ref{t:parameters}, which also shows 
the uncertainties). Note that around Hub-N there is equipartition between the perturbed 
(turbulent) and ordered magnetic field 
energies whereas around Hub-S uniform magnetic field dominates energetically over turbulence. 

Finally, the CF equation can be used to derive the plane-of-sky magnetic field strength for each region (Equation~(57) of \citealt{houde2009}):

\begin{eqnarray}
\langle\,B_{\mathrm{0}}^2\rangle^{1/2}\propto\,\sigma_{v}\,n(\mathrm{H_2})^{1/2}[\langle\,B_{\mathrm{t}}^2\rangle/\langle\,B_{\mathrm{0}}^2\rangle]^{-1/2},
\label{e:b-field}
\end{eqnarray}

\noindent where $\sigma_v$ is the velocity dispersion and $n$(H$_2$) the volume density. 
The velocity dispersion was obtained from the C$^{18}$O\,(1--0) data (Busquet et al. in prep.) 
that traces the diffuse gas around the dense filaments and hubs, resulting in $\sim2$~km\,s$^{-1}$. 
It is important to point out that for the CF method, the relevant velocity dispersion
component is the one generated by turbulence in the ISM.
For molecular clouds, the thermal velocity dispersions are typically much smaller than the 
non-thermal velocity dispersions, so it is reasonable to assume that 
$\sigma_{v} = \sigma_{v(\mathrm{NT})}+\sigma_{v(\mathrm{thermal})} \approx \sigma_{v(\mathrm{NT})}$.
Moreover, the non-thermal velocity dispersion component can be produced by turbulent motions,
gravitational infall, or rotation. Although numerous star-forming 
regions present signatures of infall even at larger scales \citep{2013peretto, 2014peretto, 
2014duarte-cabral, 2014henshaw, 2015liubaobab, 2016campbell, 2016wyrowski}, 
it is unclear whether it would cause a significant effect in the 
observed line-widths in comparison with turbulence, specially in the diffuse regions
around filaments. In this work, we assume that the velocity dispersion
derived from the C$^{18}$O\,(1--0) data is mostly due to turbulent motions, but this
is a matter that will require further investigation.

The ordered large-scale magnetic field strength component in the plane of the sky, 
$\langle\,B_{\mathrm{0}}^2\rangle^{1/2}$, for each defined region is listed in Table~\ref{t:parameters},
where the uncertainties are derived from the range of volume densities $n$(H$_2$). Considering the 
entire set of H-band polarization data associated with IRDC G14.2 the sky-projected magnetic field 
strength component is  0.39~mG, while for Hub-N and Hub-S they are given by $\sim$0.32 and 0.55~mG, respectively. 

It is important to point out that if the total magnetic field $B_{\mathrm{tot}}$ has an inclination 
$\beta\neq90\degr$ with respect to the line-of-sight, then the CF calculation 
will lead to underestimated values, since what is being measured is only the 
plane-of-sky component: $B_{\mathrm{pos}}=B_{\mathrm{tot}}\sin{\beta}$. 
The inclination $\beta$ is unknown and therefore it is difficult to correct for this effect 
in a precise way. However, \citet{crutcher2004} showed that it is possible to account for 
it at least statistically by integrating
over all possible $\beta$ values. That leads to the following correction which is 
being applied here: $B_{\mathrm{tot}}=(4/\pi)B_{\mathrm{pos}}$. Table \ref{t:parameters}
lists the values for the total magnetic field values computed for each region,
with its respective uncertainties.


\begin{table*}[t]
\centering
\caption{\label{tab} Physical properties in IRDC G14.2}
\begin{tabular}{lccccccc}
\tableline\tableline
Region 		&$\langle\,B_{\mathrm{t}}^2\rangle/\langle\,B_{\mathrm{0}}^2\rangle$   &$B_{\mathrm{pos}}$ (mG)   & $B_{\mathrm{tot}}$ (mG) &$M\,(\mathrm{M}_{\odot}$)	&$N_{\mathrm{H_2}}$ (cm$^{-2}$) & $\lambda$ & $M_A$ \\ \hline
Cloud    	&$0.86\pm0.62$	&$0.31^{+0.28}_{-0.07}$	&$0.39^{+0.36}_{-0.09}$			&4660		&$2.8\times10^{22}$		&0.6		&0.7	 \\
     Hub-N  	&$1.33\pm0.95$	&$0.25^{+0.22}_{-0.06}$	&$0.32^{+0.28}_{-0.08}$ 		&2000		&$4.5\times10^{22}$		&1.1 		&0.8	 \\
     Hub-S  	&$0.46\pm0.33$  &$0.43^{+0.38}_{-0.10}$ &$0.55^{+0.48}_{-0.13}$			&1550		&$3.5\times10^{22}$		&0.5		&0.5	 \\
     
\hline
\end{tabular}
\tablecomments{Following \citet{houde2009} method, the table respectively lists for each defined region the turbulent to uniform magnetic energy ratio, $\langle\,B_{\mathrm{t}}^2\rangle/\langle\,B_{\mathrm{0}}^2\rangle$, the magnetic field strength in the plane of the sky, $B_{\mathrm{pos}}$, derived using the CF relation (Eq~\ref{e:b-field}), the total magnetic field, $B_{\mathrm{tot}}$, the mass, the column density, the mass-to-magnetic-flux ratio ($\lambda$) and the Alfv\'en Mach number ($M_A$).
}
\label{t:parameters}
\end{table*}

\subsection{Estimates of mass-to-magnetic-flux ratios and Alfv\'en Mach numbers}

To understand if magnetic fields are strong enough to support clouds against gravitational 
collapse, it is useful to study the mass-to-magnetic-flux ratio ($M/\Phi$) which is 
conveniently calculated relative to a critical value given by 
$(M/\Phi)_{\mathrm{crit}}=1/2\pi\sqrt{G}$ \citep{nakano1978}, where $G$ is the gravitational constant 
and $\Phi$ is the magnetic flux. \citet{crutcher2004} showed that this relative quantity
may be expressed as a function of the $\mathrm{H_{2}}$ column density ($N_{\mathrm{H_{2}}}$)
and the total magnetic field strength:

\begin{equation}
\lambda = \frac{(M/\Phi)}{(M/\Phi)_{\mathrm{crit}}} = 7.6 \times 10^{-21}  \Big(\frac{N_{\mathrm{H_{2}}}}{\mathrm{cm}^{-2}}\Big)\Big(\frac{B_{\mathrm{tot}}}{\mu\mathrm{G}}\Big)^{-1}
\end{equation}

\noindent It is known that $\lambda$ can be affected by the geometry of the cloud \citep{crutcher2004}.
However, given the intricate arrangement of filamentary features at the IRDC G14.2 region, we 
chose not to make any assumptions regarding its morphology.



Furthermore, in order to access the importance of the interstellar turbulent motion in disturbing the 
magnetic field lines, we calculate the Alfv\'en Mach number, which is given
by:

\begin{equation} 
M_{\mathrm{A}}=\frac{\sqrt 3\, \sigma_{v}}{V_{A}}
\end{equation}

\noindent
where $V_{A}=B_{\mathrm{tot}}/\sqrt{4\pi \rho}$ is the Alfv\'en speed.
$M_{\mathrm{A}}$ can be viewed as a measure of the ratio between the turbulent 
and magnetic energies (in fact, this ratio is given by $M_{\mathrm{A}}^{2}$), 
and therefore the sub-alfv\'enic ($M_{\mathrm{A}} < 1$) or super-alfv\'enic ($M_{\mathrm{A}} > 1$) 
conditions indicate whether the relative importance of magnetic field support against the 
gravitational collapse is higher or lower as compared to turbulence in the ISM.
Notice that similarly to the CF method, we assume that the non-thermal motions
are dominated by turbulence.

To obtain the mass and column density of each defined region we integrate the dust continuum emission 
at 870~$\mu$m \citep{busquet2010} over the same area where  $B_{\mathrm{pos}}$ is measured. 
Notice that this integration also includes the dense structures within the selected areas, since the goal 
of calculating $\lambda$ is to evaluate the gravitational stability of the cloud against magnetic field support.
In cold and dense clouds like IRDCs dust grains are supposed to be coagulated and covered of icy mantles \citep{2009peretto}, so we derived the mass by adopting a dust mass opacity coefficient at 870~$\mu$m of 1.7~cm$^2$\,g$^{-1}$, which corresponds to agglomerated grains with thick ice mantles in cores of densities $\sim10^5$~cm$^{-3}$ \citep{ossen1994}, and assuming that the dust emission at 870~$\mu$m is optically thin, a gas-to-dust ratio of 100, and a dust temperature of 17~K. Dust temperature has been obtained using the rotational temperature derived from NH$_3$ data of Paper~I and converted to kinetic temperature though the prescription adopted by \citet{tafalla2004}. For the column density, $N(\mathrm{H}_2)=M/\mu\,m_{\mathrm{H}}A$, where $\mu=2.8$ is the molecular weight per hydrogen molecule, $m_{\mathrm{H}}$ is the mass of the hydrogen, and $A$ is the area used to derive the mass. The final values of $M$, $N(\mathrm{H_2})$, 
$\lambda$, and $M_{\mathrm{A}}$ are reported in Table~\ref{t:parameters}. 
As with the magnetic field values, the uncertainties in $\lambda$ and $M_{\mathrm{A}}$ can reach around a factor of 2.
Similar values of $M_{\mathrm{A}}$ and $\lambda$ are found by \citet{pillai2015} toward two massive IRDCs using submillimeter polarization data.

\section{Discussion}
\label{s:discussion}

\subsection{Cloud and filament formation through gravitational collapse parallel to magnetic field lines}
\label{disc_filform}

The polarization data from large to small scales at the IRDC G14.2 region show that 
not only magnetic fields are tightly perpendicular to the star-forming
dense filamentary structures within (with a few exceptions, as discussed below), 
but also the cloud as a whole (in which the 
filaments constitute the densest parts at the center) has a long-shaped morphology
perpendicular to the local magnetic field lines. This suggests a scenario 
in which magnetic fields have played an important role in regulating the gravitational 
collapse, being dynamically important in shaping elongated ISM structures from 
size scales of $\sim30$ pc down to $\sim2$ pc.

It is obvious from Figure \ref{compcloud} that there is a large dispersion in the 
relative orientation between the R-band segments and the cloud. This is not surprising, 
given that there are numerous hierarchical substructures and diffuse filamentary features 
around the entire region, as shown by the {\it Herschel} image (Figure \ref{polmap}c).
Some coupling between the magnetic field lines and these diffuse clouds are expected, 
which may explain a fraction of the dispersion observed. However, the general trend 
of magnetic fields perpendicular to the cloud is still evident.

At smaller scales ($\sim2\,$pc), the analysis on Figure \ref{compfilaments} shows that filaments 
and hubs are remarkably well oriented perpendicularly to magnetic field lines.
It is interesting to see that field lines show some smooth variations in orientation
inside this area, and the orientation of filaments seem to follow these smooth variations. 
This is a further indication that magnetic fields favored the gravitational collapse of these
structures parallel to field lines.

Two important exceptions are: Hub-N, which exhibits a slight trend of magnetic field lines
parallel to the structure; and F10-E, which shows only a marginal perpendicular correlation 
with the filament axis. 
It is possible that the original field
morphology in this area has been disrupted due to its proximity with 
IRAS$\,$18153-1651, an ultra-compact H{\sc ii} region seen in the {\it Spitzer} $8\,\mu$m image 
as a bright extended area right next to Hub-N (Figure \ref{compfilaments}a).
Paper I showed that this hub has likely been heated by the interaction with 
the ultra-compact H{\sc ii} region, and its NH$_{3}$ velocity is consistent with an expanding shell.
This is consistent with the fact that the turbulent-to-uniform magnetic 
field energy ratio ($\langle\,B_{\mathrm{t}}^2\rangle/\langle\,B_{\mathrm{0}}^2\rangle$)
is higher in Hub-N, compared to Hub-S and the entire cloud.

Recent observations show that the presence of magnetic fields aligned perpendicularly to filaments seems to be an ubiquitous characteristic of star-forming clouds \citep[e.g.,][]{franco2010,li2013,zhang2014}, at least when considering densities above a certain threshold. 
The most recent evidence
comes from the all-sky polarimetic observations of the {\it Planck} space telescope: by analysing a group of 
nearby molecular clouds, \citet{2015soler} showed that the relative orientations studied
as a function of column density gradually changes from preferentially parallel or random 
to preferentially perpendicular. Furthermore, previous works by \citet{goldsmith2008}
and \citet{tassis2009} also showed that within dense environments, magnetic fields
are most likely perpendicular to the main filamentary structures, perhaps even being 
responsible for channeling interstellar material through diffuse striated features 
also perpendicular to the filaments. More recently, \citet{zhang2014} surveyed a sample of 14 massive star forming clumps and filaments at 870~$\mu$m using the polarimeter on the Submillimeter Array. By comparing the dust polarization at dense core scales of $0.01-0.1$~pc with the pc-scale polarization, they concluded that magnetic fields play an important role in channeling gas during the collapse of the clump and the formation of dense cores. Therefore, magnetic fields appear to be dynamically important even at scales smaller than 1~pc.

Particularly in IRDC G14.2, Paper I pointed out that some striations are seen in the 
NH$_{3}$ map, converging towards filament F10-E. A visual inspection of the H-band polarization 
map shows that segments superposed to the striations are parallel to them, and perpendicular 
to the main filament, suggesting flows of material possibly converging into the main filament are parallel
to magnetic fields (red arrows in Figures \ref{polmap}b and \ref{polmap}d). 
Some striations parallel to polarization segments may also be seen after 
a close visual inspection of the H$\alpha$ image (Figure \ref{polmap}b, red arrows along its bottom-left
portion), identified as dark patches observed against a bright extended emission.
This suggests a scenario similar to the ones observed in the Taurus molecular cloud \citep{goldsmith2008},
in the Riegel-Crutcher cloud \citep{mcclure2006}, and in Lupus I \citep{2015franco}. However, in these three 
examples, the interstellar structures were nearby, which allowed a clearer view of the diffuse striations.

It is instructive to point out that an alternate explanation for the perpendicular condition between
filaments and magnetic field lines could be proposed: the same configuration would be expected if
magnetic field lines were dragged inwards by infalling material, which could also produce
the striations previously mentioned. However, it is difficult to reconcile this scenario with the fact that
magnetic fields at large-scales are also perpendicular to the filamentary features inside the cloud.
In addition, the magnetically
dominated gravitational collapse scenario is supported by MHD simulations, as described in Section \ref{s:models}.

\subsection{Comparison with simulations and analysis of stability against magnetic field 
support and turbulent motions}
\label{s:models}

Recently, \citet{vanloo2014} developed numerical simulations designed to model the non-linear 
evolution of a gravitational instability within a layer of interstellar material threaded 
by magnetic fields. The simulations show that although the presence of magnetic fields 
doesn't seem to influence on settling the filaments' central density profiles (which 
is more consistent with a typical hydrodynamical equilibrium structure), they play 
an important role in determining their morphological and spatial distribution. While 
weak magnetic fields lead to spiderweb-like filamentary features, strong magnetic fields
often generate a network of parallel filaments aligned perpendicular to field lines.

Given the similarities of the model outcomes with the morphological features of IRDC G14.2,
\citet{vanloo2014} compared their simulations with 
a fraction of the IRDC G14.2 area (specifically around Hub-N) using the polarimetric data from 
this work that was available at that time in Paper I. They find that the 
formation of these filaments is consistent with fragmentations of a layer threaded 
with strong magnetic fields, leading to parallel elongated structures perpendicular to 
field lines. The polarimetric observations from the present work provides further support 
for this model, and generalizes its conclusions for the entire filamentary network of
IRDC G14.2. The high magnetic field strengths estimated here ($\approx320-550~\mu$G) support 
a scenario in which the initial conditions favored a collapse of density perturbations 
parallel to magnetic fields, leading to the morphology of parallel filaments 
currently observed. \citet{vanloo2014} estimated that for IRDC G14.2, the magnetic field
values would need to be stronger than $12-25~\mu$G in their ``strong magnetic field" model, 
in which parallel filaments are expected to be formed. Our estimated values are one 
order of magnitude higher than this lower limit, showing that IRDC G14.2 is well into 
the strong magnetic field regime.

Alfv\'en Mach numbers ($M_{\mathrm{A}}$) calculated for each defined region show that the sub-alfv\'enic condition
is pervasive at these small scales, implying that the magnetic field strength dominates 
over the turbulent motion. Furthermore, the values of $\lambda$ are in the range $0.5 - 1.1$, 
suggesting a sub-critical condition (although they are close to the critical value, especially
considering that there is an uncertainty in the cloud's thickness). 
However, active star formation is already taking place \citep{wang2006,povich2010}, suggesting that
although magnetic fields seem to be strong enough to dominate over turbulence, 
it was usually not sufficient to prevent the gravitational collapse, which eventually 
led to star formation. Therefore the close-to-critical condition might be related to the 
filaments' envelopes, while the denser interior (not probed by the polarization data) has probably
reached supercritical conditions. $\lambda$ values may depend on whether the envelopes or the 
cores are probed \citep{bertram2012}.

\subsection{Magnetic fields related to the evolutionary sequence of the IRDC G14.2 complex}

Using single-dish $^{12}$CO observations, \citet{elm1976} provided the first description 
of the molecular cloud in which IRDC G14.2 is located, dividing the region into four 
fragments named A-D. Fragment C is roughly coincident with the position of IRDC G14.2. 
According to \citet{elm1976}, these fragments seem to be part of an evolutionary sequence:
nearby star-forming region M17, together with fragments A and B, are somewhat more evolved, 
while fragments C and D appear to be younger. 

Using the densities and velocity dispersions from the $^{12}$CO data, \citet{elm1976} 
estimated that the fragments appear to be contracting on a time scale
which is 2-3 times larger than the free-fall time, suggesting that strong internal magnetic fields
of $\sim340\,\mu$G could be providing some support against the collapse in fragment C. 
Their estimate, which is based on equipartition is remarkably 
similar to the values of magnetic field strengths computed in this work for the filamentary 
structures within IRDC G14.2.
However, it is important to point out that interstellar structures with larger
aspect ratios (such as filamentary features) have longer collapse timescales as compared
to spherical clouds \citep{2012pon}. Thus, an alternative explanation for the 2-3 times discrepancy
in contraction time observed by \citet{elm1976} is due to the filamentary nature of the cloud, 
which couldn't be inferred using the low resolution $^{12}$CO data.

Another interesting evolutionary aspect of this region, revealed by \citet{povich2010}, 
is that there seems to be a lack of O-type stars, leading to an initial mass function
significantly steeper than the Salpeter IMF. It is unclear, however, whether the 
support against gravitational collapse provided by strong magnetic fields, 
had any influence on halting or delaying the formation of massive stars.

\subsection{Magnetic fields at the RCW 157 H{\sc ii} region}
\label{sec_rcw157}

In mapping the large scale interstellar polarization around IRDC G14.2, a significant fraction of the 
RCW 157 H{\sc ii} region was covered (top-right of Figure \ref{polmap}a and c). Therefore, as a side-product of this work, it
offers the opportunity to analyze the magnetic field morphology in this structure at least
in a qualitative manner. Figure \ref{polmap}a shows that this area is dominated by a bright 
H$\alpha$ extended emission. Pillars and ``elephant trunks" are seen as dark patches 
in absorption against this bright H$\alpha$ glow, 
extending inwards at the edge of the H{\sc ii} region (black arrow in Figure \ref{polmap}a). 
These finger-shaped features are usually
generated by radiatively-driven effects, and are commonly observed in this kind of environment.

It is clear that the general polarization orientation towards RCW 157 is markedly different from the 
southern areas (compare Figure \ref{polmap}c above and below the dashed yellow line): the segments
usually span orientations between 80 and 100$\degr$, while the typical large-scale orientation 
in the IRDC G14.2 area is $\approx 140\degr$. Moreover, although several interstellar substructures are 
observed at RCW 157, the magnetic field morphology seems fairly well oriented: particularly at the 
northern portion of the map ($\alpha < 18^{\mathrm{h}}17^{\mathrm{m}}30^{\mathrm{s}}$ and $\delta > -16\degr38'$),
the angular dispersion is only $15\degr$. Furthermore, along the edges of the H{\sc ii} region,
polarization segments in general are parallel to the borders (i.e., parallel to the dashed
yellow line). In previous works, it has been shown that the expansion of an H{\sc ii} region
can modify the original magnetic field orientation, pilling up field lines along 
its borders \citep{santos2012,santos2014}. The higher magnetic field strength due to
the pilling effect can lead to low polarization angle dispersions.
These qualitative features observed at RCW 157 suggest that a similar effect might be ongoing
in this area. The uniformly-oriented polarization segments are probably probing 
the expanding interstellar shell along the line-of-sight.

It is also interesting to see that the finger-shaped pillars are parallel to polarization 
segments. This configuration is expected, because during the formation of these structures, 
magnetic fields are swept out by the expanding front and its lines are wrapped around the pillars.
These observations give support to radiation-MHD simulations of H{\sc ii} regions forming within
magnetized molecular clouds, which predict very similar characteristics \citep{peters2011,arthur2011}.

\section{Conclusions}
\label{s:conclusions}

In this work we have studied the morphological relation between 
magnetic fields and the various interstellar structures at the 
IRDC G14.2 star-forming complex. Our goal was achieved through polarimetric observations 
of background stars in the optical and NIR spectral bands, 
aimed respectively at the large-scale cloud and the small-scale
filamentary structures within its densest portions. The analysis was 
carried out after careful removal and correction of the foreground polarization 
component. Below is a list of the main conclusions:

\begin{enumerate}
\item We compared the orientation 
of magnetic fields with filaments and hubs, and also with the molecular cloud 
in which these structures are embedded. It is clear that 
magnetic fields are perpendicular both to the small-scale filamentary features
and to the large-scale cloud. For filaments, this condition holds true with few exceptions even 
when considering Monte Carlo simulations which account for sky-projection effects. 
These characteristics are consistent with a scenario in which magnetic fields
regulated the gravitational collapse from large ($\approx 30\,$pc) to small 
scales ($\approx 2\,$pc);

\item Combining the polarization data with dust emission and molecular line
observations, we estimate total magnetic fields strengths,
Alfv\'en Mach numbers and mass-to-magnetic-flux ratios. 
The structures are predominantly in a sub-alf\'enic and in close-to-critical condition, 
suggesting that magnetic fields are strong enough to overcome turbulent
motions, but not sufficient to prevent the gravitational collapse. 
The high magnetic field values corroborate previous numerical simulations 
that show that these conditions eventually lead to a gravitational instability 
developing along magnetic field lines, therefore generating filaments 
organized in a parallel arrangement;

\item The range of magnetic field values obtained for the filaments and hubs 
($\approx 320 - 550\,\mu$G) is consistent with estimates based 
on simple equipartition assumptions by \citet{elm1976}, who suggested that internal magnetic 
field strengths would be around $340\,\mu$G. According to their interpretation, 
the presence of such strong magnetic fields might be a necessary condition to explain why
the large-scale cloud is possibly contracting in a time scale $2-3$ times
larger than what expected from the free-fall time.
\end{enumerate}

As a precursor to a massive OB association presenting numerous filamentary
interstellar features and young stellar sources, the IRDC G14.2 cloud proves to be an
ideal star-forming site to study the underlying physical conditions regulating the 
gravitational collapse. This is an important target for additional analysis, 
particularly using high-resolution polarization emission surveys (in the far-infrared
or submillimeter wavelengths) or even spectral data focused on Zeeman splitting. 
This would be a natural continuation of this work, 
given the significant role played by magnetic fields in shaping the filamentary 
morphology and regulating the collapse.
More specifically, magnetic field strengths (along with $M_{\mathrm{A}}$ and 
$\lambda$ values) could be better constrained with this kind of observation, 
specially if comparisons with numerical simulations are made, assuming the specific 
physical conditions of this cloud and its sub-structures.

\acknowledgements
We are grateful to the anonymous referee for the valuable suggestions and comments.
We thank the staff of OPD/LNA (Brazil) for their hospitality and invaluable help during our observing runs. 
G.A.P.F. and F.P.S. acknowledge support from the Brazilian agencies CNPq, CAPES and FAPEMIG.
F.P.S. was supported by the CAPES grant 2397/13-7. G.B. acknowledges the support from the Spanish Ministerio de Economia y Competitividad (MINECO) under grant FPDI-2013-18204. G.B. and J.M.G. are supported by the Spanish MICINN grant AYA2011-30228-C03 and the MINECO grant AYA2014-57369-C3. 
This investigation made extensive use of data products from the Two Micron All Sky Survey (2MASS), 
which is a joint project of the University of Massachusetts and the Infrared Processing and 
Analysis Center/California Institute of Technology, funded by the National Aeronautics and 
Space Administration and the National Science Foundation.  
This research is based in part on observations made with the {\it Spitzer} Space Telescope, which 
is operated by the Jet Propulsion Laboratory, California Institute of Technology under a 
contract with NASA.
We are grateful to Drs. A. M. Magalh\~aes and A. Pereyra for providing the polarimetric unit 
and the software used for data reductions. 

{\it Facilities:} 
\facility{LNA: 1.6\,m}

\bibliography{astroref}
\bibliographystyle{aa}

\end{document}